\RequirePackage{fix-cm}
\RequirePackage{silence}
\documentclass[fleqn,usenatbib,useAMS]{mnras} 

\usepackage{newtxtext,newtxmath}

\usepackage{graphicx}
\usepackage{amsmath}

\usepackage{amsfonts}
\usepackage{float}
\usepackage{bm}
\usepackage{ae,aecompl}
\usepackage{array}
\usepackage{soul}
\usepackage{mathtools}
\usepackage{multirow}
\usepackage[T1]{fontenc}
\usepackage[utf8]{inputenc}
\usepackage{booktabs}
\usepackage{graphicx}
\usepackage{subcaption}
\usepackage[normalem]{ulem}
\usepackage{subfiles}

\DeclareRobustCommand{\VAN}[3]{#2}
\let\VANthebibliography\thebibliography
\def\thebibliography{\DeclareRobustCommand{\VAN}[3]{##3}\VANthebibliography}

\DeclareRobustCommand{\appropto}{\mathrel{\vcenter{
		\offinterlineskip\halign{\hfil$##$\cr 
			\propto\cr\noalign{\kern2pt}\sim\cr\noalign{\kern-2pt}}}}}

\defcitealias{Haslbauer_2020}{HBK20}

\title[MOND cosmological simulations on Gpc scales]{How does a MOND cosmology fare on Gpc scales? $-$ Collisionless $N$-body simulations of $\nu$HDM} 

\author[A. Russell et al.]{Alfie Russell$^{1}$\thanks{E-mail: \href{mailto:apr8@st-andrews.ac.uk}{apr8@st-andrews.ac.uk} (Alfie Russell); \newline \hspace*{4em} \href{mailto:indranil.banik@port.ac.uk}{Indranil.Banik@port.ac.uk} (Indranil Banik); \newline \hspace*{4em} \href{mailto:hz4@st-andrews.ac.uk}{hz4@st-andrews.ac.uk} (Hongsheng Zhao)}, Indranil Banik$^{2, 1}$, Oscar Cray$^{1}$ and Hongsheng Zhao$^{1, 3, 4, 5}$ \vspace{10pt} \\
$^{1}$Scottish Universities Physics Alliance, University of Saint Andrews, North Haugh, Saint Andrews, Fife, KY16 9SS, UK \\ 
$^{2}$Institute of Cosmology and Gravitation, University of Portsmouth, Dennis Sciama Building, Burnaby Road, Portsmouth PO1 3FX, UK \\
$^{3}$Department of Astronomy, School of Physical Sciences, University of Science and Technology of China, Hefei, Anhui 230026, China \\
$^{4}$Institute for Advanced Study in Physics, Zhejiang University, Hangzhou, 310058, China \\
$^{5}$School of Physics, Zhejiang University, Hangzhou, 310058, China}

\date{Accepted XXX. Received YYY; in original form ZZZ}

\pubyear{\the\year{}}

\begin{document}
\label{firstpage}
\pagerange{\pageref{firstpage}--\pageref{lastpage}}
\maketitle

\begin{abstract} 
We present the largest collisionless $N$-body cosmological simulations in a MOdified Newtonian Dynamics (MOND) cosmology to date. Our 4 simulations cover $\Lambda$CDM as a baseline, a MOND with hot dark matter model known as $\nu$HDM, and 2 unphysical models we call $\Lambda$HDM and $\nu$CDM to test the individual contributions of hot dark matter and MOND gravity, respectively. $\nu$HDM reproduces the CMB power spectrum while also theoretically matching cluster dynamics and preserving MOND predictions for galactic rotation curves. We test its viability on cosmological scales using simulations with $256^{3}$ particles in a box of size $800/h$ comoving Mpc. We find generically that the MOND models massively overproduce large-scale structures by $z=0$, with a most massive cluster in $\nu$HDM of $\approx 5 \times 10^{17} \, M_{\odot}/h$ and typical peculiar velocities of several thousand km/s. We also explore a local void solution to the Hubble tension in these models. Analogues to the observed ``Local Hole'' do form in the MOND models, but values for the deceleration parameter $<-1.5$ in these regions prevent a satisfactory resolution to the Hubble tension. While $\Lambda$CDM significantly underpredicts the observed bulk flow in Cosmicflows-4, the high peculiar velocities that arise in the MOND models create the opposite problem, ruling out $\nu$HDM at $>5\sigma$ confidence. Observations clearly require a much milder enhancement to the rate of structure growth in $\Lambda$CDM than is provided by the $\nu$HDM paradigm. Our results also suggest that replacing cold dark matter with hot dark matter is unlikely to provide a viable cosmological model, regardless of the gravity law.

\end{abstract}


\section{Introduction}
\label{sec:Introduction}

A major accomplishment of modern cosmology is an understanding of how structure in the universe grew from small density fluctuations caused by quantum uncertainty in the era of cosmic inflation at very early times \citep{Guth_1981}. These fluctuations oscillated in strength prior to recombination due to tight coupling between the baryons and photons at that time, a consequence of the high Thomson scattering cross-section of free electrons. Modes of different wavelengths had different oscillation frequencies, but they all oscillated for a fixed time interval of about 380~kyr between the Big Bang and recombination. This created a characteristic pattern of oscillations in the power spectrum of the Cosmic Microwave Background \citep[CMB;][]{Bernardis_2000, WMAP_2013, Planck_2020, ACT_2025}.

In a universe governed by General Relativity \citep[GR;][]{Einstein_1915} with only baryonic matter, the CMB power spectrum would have a very simple form governed purely by diffusion damping, causing each peak to be lower in amplitude than the previous peak by a similar factor \citep{Hu_2002}. This is contrary to the observation that the third peak has a similar amplitude to the second peak, but both have a much lower amplitude than the first peak, violating the expectation that the peak amplitudes decrease in a continuous manner \citep{Spergel_2003, Spergel_2007}.

Another unusual aspect of the CMB is that once we allow for a dipole anisotropy due to the motion of the Sun relative to the CMB \citep{Kogut_1993}, the remaining fluctuations are at the level of $10^{-5}$. In linear structure formation theory, the fractional density perturbation $\delta \propto a$ in the era of matter domination, where $a$ is the cosmic scale factor normalized to unity today. The assumption of matter domination is quite accurate because recombination occurred when the universe was $\ga 7 \times$ older than at the epoch of matter-radiation equality \citep{Banik_2025_cosmology}, while dark energy has only recently become important. Including it only worsens the problem as it slows down structure growth, as does radiation \citep{Meszaros_1974}. The temperature fluctuations on the small scales relevant to galaxy formation are smaller still due to diffusion or Silk damping \citep{Silk_1968}. Since recombination occurred at redshift $z = 1090$, it is clear that an additional ingredient is required to form gravitationally bound structures on galaxy scales by the present epoch.

In the $\Lambda$-Cold Dark Matter ($\Lambda$CDM) standard model of cosmology \citep*{Efstathiou_1990, Ostriker_Steinhardt_1995}, the missing ingredient in structure formation theory is CDM. Density perturbations in the CDM could grow as soon as the universe became dominated by matter. The lack of any coupling to radiation allowed these perturbations to grow much more than perturbations in the baryons, which were still very small by the time of recombination. 

Besides the issue of structure formation, CDM is also necessary to explain the internal dynamics and stability of many gravitationally bound objects in the local universe. Self-gravitating discs of stars are unstable \citep{Miller_1968, Hockney_1969, Hohl_1971}, leading to the suggestion of massive dark haloes \citep*{Ostriker_Peebles_1973, Ostriker_Peebles_Yahil_1974}. These are also necessary to explain why M31 is currently approaching our Galaxy, even though these must have been receding from one another shortly after the Big Bang \citep[the ``timing argument'';][]{Kahn_Woltjer_1959}. More directly, the internal dynamics of these and other galaxies cannot be explained by their baryonic matter alone \citep[e.g. ][]{Rubin_1970, Rogstad_1972, Roberts_1973, Roberts_1975, Faber_1979, Mateo_1998, Sofue_2001, Tolstoy_2009, SPARC, Simon_2019, Arora_2023}. On larger scales, the dynamical masses and X-ray emissions of clusters show they are similarly more massive than their baryonic content \citep[e.g. ][]{Zwicky_1933, Biviano_2006, Sereno_2025, Cavaliere_1976, Pointecouteau_2005, Ettori_2013}. This conclusion is further reinforced by measurements of the gravitational field using gravitational lensing by both galaxies and clusters \citep[see][and references therein]{Hoekstra_2008, Natarajan_2024}. In a Universe governed by GR, a dominant CDM component is a requirement to explain astronomical observations on almost all scales beyond the Solar System.

\subsection{Milgromian dynamics (MOND)}
\label{sec:MOND}

Our preceding discussion assumes GR can be extrapolated beyond the Solar System scales for which it was originally designed. The resulting need for dark matter is concerning, as CDM particles would have exotic properties that are not observed for any known particle in the standard model of particle physics. Decades of effort with direct detection experiments have ruled out much of the parameter space originally thought to be viable \citep{LUX_2017, LUX_2023, Aalbers_2025}. Indirect detection experiments give similar conclusions based on over a decade of observations towards the supposedly CDM-dominated dwarf satellites of the Milky Way \citep{Ackerman_2015, Hoof_2020, Hu_2024_DM, Song_2024}. These results raise the possibility that CDM particles do not exist.

If disc galaxies are purely baryonic objects, then their visible mass must provide sufficient gravity to explain their observed flat rotation curves, which imply a huge missing gravity problem in the context of GR. Solving this with only the visible mass requires a modification to the dynamical laws. The most widely considered proposal along these lines is known as MOND \citep{Milgrom_1983}. The basic idea of MOND is that in an isolated spherically symmetric system with Newtonian gravity $g_{_{\mathrm{N}}}$, the actual gravity $g$ has the following asymptotic limits:
\begin{eqnarray}
    g ~\to~ \begin{cases}
    g_{_{\mathrm{N}}} \, , & \textrm{if} ~g_{_{\mathrm{N}}} \gg a_{_0} \, , \\
    \sqrt{g_{_{\mathrm{N}}} a_{_0}} \, , & \textrm{if} ~g_{_{\mathrm{N}}} \ll a_{_0} \, .
    \end{cases}
    \label{nu_cases}
\end{eqnarray}
MOND introduces $a_{_0}$ as a fundamental new constant of nature with dimensions of acceleration. The idea that gravity should be modified at low accelerations rather than beyond a fixed length was motivated by the Tully-Fisher relation between flat rotation velocity and galaxy luminosity \citep{Tully_1977}, which shows that the expected Keplerian decline to the rotation curve becomes anomalously flat at about the same $g_{_{\mathrm{N}}}$ in galaxies of different mass \citep[see sections~2 and 3.1.1 of][]{Banik_Zhao_2022}. Galaxy rotation curves require that $a_{_0} = 1.2 \times 10^{-10}$~m/s$^2$ \citep*{Begeman_1991, Gentile_2011, McGaugh_Lelli_2016}. For comparison, we do not have direct knowledge of how gravity behaves beyond the 170~AU distance to Voyager~1, our most distant currently operational spacecraft. The gravitational field strength here is $g = 2 \times 10^{-7}$~m/s\textsuperscript{2} or about $1700 \, a_{_0}$, far higher than typical accelerations in galaxies or in the early universe (see section~3.1.3 of \citealt*{Haslbauer_2020}, hereafter \citetalias{Haslbauer_2020}).

A modification to Newtonian gravity at $g_{_{\mathrm{N}}} \la a_{_0}$ is not only possible, it actually leads to unparalleled predictive success with the rotation curves of galaxies \citep{McGaugh_2020}. The MOND prediction of a tight radial acceleration relation (RAR) between $g$ and $g_{_{\mathrm{N}}}$ holds in actual data of galaxies, which show that the intrinsic scatter is $\la 0.034$~dex \citep{Desmond_2023}, with residuals from the mean RAR not showing correlations with many properties of galaxies like surface brightness and gas fraction \citep{Stiskalek_2023}. A similar amount of intrinsic scatter has been inferred using neutral hydrogen observations of a homogeneously analysed sample of 19 galaxies that probe the low-acceleration end of the RAR \citep{Varsteeanu_2025}. For comparison, \citet{Ludlow_2017} showed that the intrinsic scatter in the RAR in the EAGLE simulations of $\Lambda$CDM is 0.08~dex \citep{Crain_2015, Schaye_2015}, highlighting that it remains difficult to explain the tight RAR in this paradigm. Moreover, MOND is also successful in elliptical galaxies \citep{Lelli_2017, Chae_2020_elliptical, Shelest_2020} and low surface brightness galaxies \citep{Karukes_2017, Di_Paolo_2019, McGaugh_2021}, where enhanced tidal susceptibility arising from the non-linear behaviour of MOND provides an elegant explanation for the observed signs of disturbance of dwarf galaxies in the Fornax Cluster and the lack of low surface brightness dwarfs towards the cluster centre \citep{Asencio_2022}. Developments in stacked galaxy-galaxy weak lensing can now probe $g$ hundreds of kpc from a galaxy in a statistical sense, demonstrating that a tight RAR consistent with the rotation curve data continues to hold at much lower accelerations \citep[about 2.5~dex below $a_{_0}$ in terms of $g$ or 5~dex in $g_{_{\mathrm{N}}}$;][]{Brouwer_2021, Mistele_2024_flatRC, Mistele_2024_RAR}. Relativistic versions of MOND can explain the CMB power spectrum while having gravitational waves that travel at the speed of light \citep{Skordis_2019, Skordis_2021, Skordis_2022}. For extensive reviews of MOND, we refer the reader to \citet{Famaey_McGaugh_2012} and \citet{Banik_Zhao_2022}.

\subsection{The neutrino-Hot Dark Matter (\texorpdfstring{$\nu$}{v}HDM) model}
\label{sec:nuHDM model}

A long-standing problem with MOND is that it still faces a residual missing gravity problem in galaxy clusters \citep{Sanders_1999, Sanders_2003, Pointecouteau_2005, Eckert_2022_cluster_RAR, Li_2023, Li_2024, Kelleher_2024, Famaey_2025}. It is possible that this problem will be alleviated in a relativistic MOND theory, but this is far from guaranteed \citep{Durakovic_2024}. If we instead assume that MOND can be applied to galaxy clusters in much the same way as galaxies, we still have to assume that there is some form of undetected mass. This might be in the form of sterile neutrinos with a rest energy of 11~eV, which are the main mass component of the $\nu$HDM paradigm on cluster and larger scales \citep{Angus_2007, Angus_2009, Angus_2010, Wittenburg_2023}. The proposed sterile neutrinos are far heavier than ordinary neutrinos, whose rest energy is at most 0.45~eV \citep{Katrin_2019, Katrin_2022, Katrin_2025}. Even so, the low mass of sterile neutrinos would make them a form of HDM -- sterile neutrinos with such a low mass would barely affect the MOND fits to galaxy rotation curves \citep{Angus_2010_minimum_neutrino_mass} because their number density is limited by the Pauli Exclusion Principle \citep[the Tremaine-Gunn limit in an astrophysical context;][]{Tremaine_Gunn_1979}. However, the larger size and velocity dispersion of galaxy clusters allows for sterile neutrinos to play a greater role, so they can explain the offset between the weak lensing and X-ray peaks in the Bullet Cluster \citep{Angus_2007}. Sterile neutrinos can also account for the second and third peaks in the CMB power spectrum having a similar height, provided the sterile neutrinos have about the same average mass density as the CDM in the $\Lambda$CDM paradigm \citep*{Angus_2011_clusters, Samaras_2025}. This is indeed the case if they are in thermal equilibrium in the early universe. The sterile neutrinos would be non-relativistic at the time of recombination, acting much like CDM. The sterile neutrinos would be relativistic in the era of Big Bang nucleosynthesis (BBN), but it has been argued that the impact on the primordial deuterium and helium abundances is within observational uncertainties \citepalias[section~3.1.2 of][]{Haslbauer_2020}.

Structure formation is faster in $\nu$HDM due to the MOND gravity law \citep{Katz_2013}. This can better account for the high redshift, mass, and collision velocity of the interacting galaxy clusters known as El Gordo \citep{Menanteau_2010, Menanteau_2012}, which are incompatible with $\Lambda$CDM at $>5\sigma$ confidence for any plausible collision velocity \citep*{Asencio_2021, Asencio_2023}. The enhanced structure formation on large scales also makes it easier to form the KBC supervoid \citep*{Keenan_2013}, an extended underdensity that appears to be $46 \pm 6\%$ less dense than the cosmic mean out to 300~Mpc based on the near-infrared galaxy luminosity density in redshift space. The actual underdensity is probably around half that, but even so, the KBC void contradicts $\Lambda$CDM at $6\sigma$ confidence \citepalias[section~2 of][]{Haslbauer_2020}. Those authors estimated in their section~1.1 that if the KBC void gradually deepened with time to its presently observed underdensity, then outflows from the KBC void would cause the apparent local expansion rate to exceed the background $H_0 \equiv \dot{a}$ by about 11\%, where $H \equiv \dot{a}/a$ is the Hubble parameter at some epoch, $a$ is the cosmic scale factor normalised to unity today, overdots denote time derivatives, and 0 subscripts denote present values, so for instance $H_0$ is the present expansion rate or Hubble constant.\footnote{It is constant in space, but not over time.}

Remarkably, there is strong observational evidence for just such a mismatch in terms of the Hubble tension, the observation that the locally estimated $H_0$ is about 10\% higher than predicted in $\Lambda$CDM with parameters calibrated to the CMB power spectrum \citep{Riess_2024, Uddin_2024, Bhardwaj_2025, Freedman_2025, Jensen_2025, Said_2025, Scolnic_2025, Vogl_2025}. Observers estimate $H_0$ locally by measuring the distances and redshifts to nearby objects such as galaxies and supernovae (SNe), with the results plotted on a Hubble diagram. The redshift rises almost linearly with the distance $r$. If we assume that the redshift arises entirely from cosmic expansion, then this linear relation or Hubble law is caused by the fact that photons from an object at distance $r$ were emitted in the past, with the universe having expanded in the meantime. We can then apply the chain rule to get that $H_0 = cz'$, where primes indicate a radial derivative and $c$ is the speed of light \citep*{Mazurenko_2025}.

\citetalias{Haslbauer_2020} used semi-analytic models to consider the formation of the KBC void out of a small initial density fluctuation at $z = 9$. They found that outflow from the KBC void could quite plausibly inflate the local $cz'$ enough to solve the Hubble tension while having a lower $\dot{a}$ consistent with the CMB observations and other constraints like the ages of the oldest stars \citep[][and references therein]{Banik_2025_cosmology}. This self-consistent void outflow solution to the Hubble tension was later shown to provide a good match to baryon acoustic oscillation (BAO) measurements over the last twenty years \citep{Banik_2025_BAO} and the observed bulk flow curve, a measure of the peculiar velocity field in the local Universe \citep{Mazurenko_2024}. The observed bulk flow is around quadruple the $\Lambda$CDM expectation on the largest probed scale of $250/h$~Mpc, where $h$ is the Hubble constant in units of 100~km/s/Mpc \citep{Watkins_2023}. This leads to $>5\sigma$ tension with $\Lambda$CDM expectations, as shown in their figure~8. In short, $\nu$HDM seems to better account for El Gordo, the KBC void, and the fast observed bulk flows at $z \la 0.1$, all of which are severely problematic for $\Lambda$CDM while being based on quite different observational techniques.

However, hydrodynamical cosmological $\nu$HDM simulations indicate that galaxies only form by $z \approx 4$ \citep{Wittenburg_2023}, in strong disagreement with observations by the \emph{James Webb Space Telescope} (\emph{JWST}) of galaxies already at $z = 14$ \citep{Carniani_2024}. The stronger MOND gravity at $z \la 50$ \citepalias[section~3.1.3 of][]{Haslbauer_2020} struggles to compensate for the lack of CDM, which severely reduces the density perturbations on galaxy scales at early times \citep[see figure~1 of][]{Wittenburg_2023}. These difficulties with both paradigms suggest that both may have desirable features that should be preserved in a more complete theory. In particular, $\nu$HDM simulations can give insight into what phenomena to expect if structure formation is more rapid than expected in $\Lambda$CDM, as suggested by some observations \citep[see also][though see \citealt{Sawala_2025}]{Migkas_2021, Lopez_2022, Lopez_2024}.

In this contribution, we explore the $\nu$HDM paradigm using large collisionless simulations on Gpc scales. After covering the governing equations, we explain how we obtain the initial conditions for our $\nu$HDM simulations and how we advance them (Section~\ref{sec:Methods}). We then describe our analyses of the resulting particle phase space information (Section~\ref{sec:Analysis}). We present the results of these analyses in Section~\ref{sec:Results} and discuss our findings in Section~\ref{sec:Discussion}. We conclude in Section~\ref{sec:Conclusions}.

\section{Theory and Methods}
\label{sec:Methods}

The general methodology we follow is similar to that of \citet{Wittenburg_2023}, whose main points we briefly recap below.

\subsection{Governing equations}
\label{sec:Governing_equations}

Equation~\ref{nu_cases} is only a heuristic argument for how $g_{_{\mathrm{N}}}$ should be modified in order to explain galaxy rotation curves. It specifies only the asymptotic limits, but we can generalize it to arbitrary $g_{_{\mathrm{N}}}$ by writing that for an isolated spherically symmetric system,
\begin{eqnarray}
    g ~=~ g_{_{\mathrm{N}}} \nu \left( g_{_{\mathrm{N}}} \right),
    \label{eq:g_gN}
\end{eqnarray}
where the interpolating function $\nu$ takes argument $g_{_{\mathrm{N}}}$. In this contribution, we will use the simple interpolating function \citep{Famaey_2005}:
\begin{eqnarray}
    \nu ~=~ \frac{1}{2} + \sqrt{\frac{1}{4} + \frac{a_{_0}}{g_{_{\mathrm{N}}}}} \, .
    \label{eq:nu}
\end{eqnarray}
This introduces a fairly gradual transition between the Newtonian and deep-MOND regimes, providing a much better fit to the observations \citep[see figure~23 of][]{Banik_2024_WBT}. In particular, this simple form is numerically quite similar to the exponential form advocated by \citet*{McGaugh_Lelli_2016} based on the Spitzer Photometry and Accurate Rotation Curves \citep[SPARC;][]{SPARC} catalogue of galaxy rotation curves and near-infrared photometry:
\begin{eqnarray}
    \nu \left( \frac{g_{_{\mathrm{N}}}}{a_{_0}} \right) ~=~ \frac{1}{1 - \exp \left( -\sqrt{\frac{g_{_{\mathrm{N}}}}{a_{_0}}} \right)} \, .
    \label{eq:interpolating function SPARC}
\end{eqnarray}
Since we focus on large-scale structure in this contribution, our results are sensitive mainly to the low-acceleration limit in which $g \to \sqrt{g_{_{\mathrm{N}}} a_{_0}}$ and thus $\nu \to \sqrt{a_{_0}/g_{_{\mathrm{N}}}}$. The above exponential suppression of the MOND corrections to $g_{_{\mathrm{N}}}$ in the Newtonian regime causes some difficulties in numerical simulations, so we opt to use \autoref{eq:g_gN}, which has previously been shown to give excellent numerical performance when solved on a grid to obtain the MOND gravitational field \citep[see figure~1 of][]{Banik_2018_Centauri}.

In the quasilinear formulation of MOND \citep[QUMOND;][]{Milgrom_2010}, more complicated geometries are handled by taking the divergence of both sides in \autoref{eq:g_gN}:
\begin{eqnarray}
    \nabla \cdot \bm{g} ~=~ \nabla \cdot \left( \nu \bm{g}_{_{\mathrm{N}}} \right).
    \label{eq:div_g_gN}
\end{eqnarray}
This introduces an innovative two-step procedure in which we first have to find $\bm{g}_{_{\mathrm{N}}}$ using standard techniques based on the input density distribution, find $\nu$ using basic algebra and thereby obtain the source term for $\bm{g}$, and repeat the first step using $\nabla \cdot \left( \nu \bm{g}_{_{\mathrm{N}}} \right)$ as the density field. Only the ordinary Poisson equation has to be solved, allowing the use of standard numerical techniques, albeit applied twice. In a cosmological context, $\bm{g}_{_{\mathrm{N}}}$ is sourced only by the density less the cosmic mean density at that epoch \citep{Sanders_2001, Nusser_2002}. It has been argued that this Jeans swindle-like \citep{Jeans_1902} approach is valid in some relativistic theories of MOND \citep*{Thomas_2023_cosmology}, but this is not necessarily guaranteed.

\subsection{Models}
\label{sec:Models}

In addition to the $\nu$HDM model, we perform a benchmark $\Lambda$CDM simulation for comparison purposes. We also consider two intermediary models we dub $\Lambda$HDM and $\nu$CDM, which consist of Newtonian gravity with HDM initial conditions and MOND gravity with CDM initial conditions, respectively ($\nu$ in this context is better thought of as representing the MOND interpolating function in \autoref{eq:g_gN} rather than sterile neutrinos). We include these simulations as a reference to better understand the differences between each model, since directly comparing $\Lambda$CDM to $\nu$HDM makes it difficult to distinguish whether any differences in behaviour are due to a lack of CDM or due to the MOND gravity law. The different types of model we consider are summarized in Table~\ref{tab:Model_summary}. We note that although one can envisage other approaches to MOND on cosmological scales with the same limiting behaviour in galaxies, a lack of CDM is a generic prediction of any MOND cosmology.

\begin{table}
    \centering
    \begin{tabular}{lcc}
        \hline
        & \multicolumn{2}{c}{Dominant matter component} \\
        Gravity law & CDM & HDM \\ \hline
        Newton & $\Lambda$CDM & $\Lambda$HDM \\
        MOND & $\nu$CDM & $\nu$HDM \\ \hline
    \end{tabular}
    \caption{Summary of the models considered in this contribution.}
    \label{tab:Model_summary}
\end{table}

\subsection{Initial conditions}
\label{sec:ICs}

The initial conditions for our simulations were generated using the Code for Anisotropies in the Microwave Background \citep*[\textsc{camb};][]{Lewis_2000} and MUlti Scale Initial Conditions for cosmological simulations \citep[\textsc{music};][]{Hahn_2011}. A basic overview of the process is that \textsc{camb} generates a power spectrum $P(k)$, which \textsc{music} then samples to produce the initial positions and velocities of the simulation particles. While \textsc{music} does have an inbuilt functionality to produce its own spatial matter power spectrum, \textsc{camb} was used instead due to its increased flexibility with user defined cosmological parameters.

The first step is to select the cosmological parameters that should be used by \textsc{camb}. Since an explicit assumption of our cosmological MOND models is that the standard $\Lambda$CDM interpretation of the CMB temperature anisotropies is correct, we set most of our cosmological parameters equal to the values found by \citealt{Planck_2020} (see the sixth column of their table~2). We therefore adopt the following parameters: $\Omega_{\mathrm{m}} = 0.314595$, $\Omega_{b} = 0.049271$, $\Omega_{\Lambda} = 0.685405$, $\sigma_{8} = 0.8101$, $n_{s} = 0.9660499$, and $H_0 = 67.4$ $\mathrm{km}$ $\mathrm{s^{-1}}$ $\mathrm{Mpc^{-1}}$ ($h = 0.674$). The density parameters for CDM and neutrinos ($\Omega_c$ and $\Omega_{\nu}$, respectively) depend on the model as follows:
\begin{eqnarray}
    \label{eq:CDM denisty params}
    \Omega_{c} = 
    \begin{cases}
        0.264157 & \text{(CDM)}, \\
        0 & \text{(HDM)}
    \end{cases}
\end{eqnarray}
and
\begin{eqnarray}
    \Omega_{\nu} = 
    \begin{cases}
        0.001167 & \text{(CDM)}, \\
        0.265324 & \text{(HDM)}.
    \end{cases}
\end{eqnarray}
In the HDM simulations, we remove the CDM component and fill up the remaining matter budget with an extra species of sterile neutrino. These can be incorporated into \textsc{camb} by modifying the `.ini' file \citep[as laid out by][]{Angus_2009, Wittenburg_2023} to include the following parameters:
\begin{multline}
    \texttt{massless\_neutrinos} = 2.0293 \notag \\
    \texttt{nu\_mass\_eigenstates} = 2 \notag \\
    \texttt{massive\_neutrinos} = 1 \; 1 \notag \\
    \texttt{nu\_mass\_degeneracies} = 1.0147 \; 1 \notag \\
    \texttt{nu\_mass\_fractions} = 0.0044 \; 0.9956. \notag \\
\end{multline}
This ensures that the properties of the active neutrinos are unchanged, while also adding in our sterile neutrino component. There is no explicit input for the sterile neutrino mass here. Instead, \textsc{camb} assumes the neutrinos are thermalized at recombination, which then leads to a self-consistent calculation of the particle mass, this being $\approx 11$~eV/$c^2$ for this choice of parameters. A limitation here is that \textsc{camb} is not designed to implement heavier sterile neutrino masses as they will not be thermal in origin (i.e., their relic abundance would not match the active neutrino relic abundance at recombination). An analytic cut-off to the CDM power spectrum \citep[such as that described by][]{Abazajian_2006} would suffice if heavier sterile neutrinos were desired, but we want the total mass density to remain the same, requiring a lower number density. Results in this case would presumably be intermediate between our MOND simulations with CDM and HDM, i.e., our $\nu$CDM and $\nu$HDM models, respectively (Table~\ref{tab:Model_summary}).

The next step is to define an initial power spectrum $P_i(k)$ and evolve the corresponding initial density perturbations to the starting simulation redshift of 199 ($a = 0.005$). To do this, the \textsc{camb} code assumes there was an early inflationary period which produces a ``nearly scale-invariant'' power spectrum \citep{Harrison_1970, Zeldovich_1972}. This is described by
\begin{eqnarray}
    P_{i} \left( k \right) ~\propto~ k^{n_{s} - 1} \, ,
    \label{eq:Inflation power spectrum}
\end{eqnarray}
where $k$ is the comoving wavenumber and $n_{s}$ is the tilt of the scalar spectral index. The overall amplitude is determined by $\sigma_{8}$. The initial density perturbations are related to the power spectrum by
\begin{eqnarray}
    P \left( k \right) ~\propto~ \delta^{2} \left( k \right) \, .
    \label{eq:Power density relation}
\end{eqnarray}

A transfer function $T(k)$ can be calculated to relate these perturbations to those at a later redshift \citep{Eisenstein_1998}. For comparing some early redshift $z_{a}$ with a later redshift $z_{b}$, $T(k)$ would be defined as
\begin{eqnarray}
    \label{eq:Transfer definition}
    T(k) \equiv \frac{\delta \left( z_{b}, k \right)}{\delta \left( z_{a}, k \right)} \, ,
\end{eqnarray}
which can be computed numerically using linear perturbation theory. One caveat is that density perturbations at both $z_{a}$ and $z_{b}$ must still be in the linear regime, which will limit the starting redshift of the simulation.

From Equations~\ref{eq:Power density relation} and \ref{eq:Transfer definition}, we can see that $P(k)$ and $T(k)$ can be directly related by
\begin{eqnarray}
    P_{b} \left( k \right) ~=~ P_{a}(k) \, T^{2} \left( k \right) \, ,
    \label{eq:Power Transfer relation}
\end{eqnarray}
where we use $a$ and $b$ subscripts to denote the two redshifts we are considering. Therefore, once the transfer function has been calculated by \textsc{camb}, it is used to calculate the power spectrum at the starting redshift of the simulation for use by \textsc{music}. 

For these simulations, the early redshift $z_a$ would occur just after the end of the inflationary period and be determined by \textsc{camb}, while $z_b$ would correspond to when we start our simulations. We select the starting redshift as $z_{i} = 199$. This is late enough that the radiation component of the universe (which will not be simulated) has a negligible density, while also early enough that the MOND effects are suppressed. This second point is justified by the work of \citet{Haslbauer_2020}, who find that density perturbations do not enter the MOND regime until $z \la 50$. We will further corroborate this prediction in Section~\ref{sec:pec vel}.

In CDM models, the transfer function is always positive at epochs much later than recombination because the CDM coalesces in the initial overdensities, thereby preserving them during the oscillations of the baryon-photon fluid. The same logic applies to underdensities. Initial overdensities do not turn into underdensities, and vice versa. Baryons do of course undergo oscillations (Section~\ref{sec:Introduction}), but they subsequently fall into the much deeper potential wells generated by the CDM, so the initial density perturbation has the same sign as the density perturbation subsequent to recombination after a very short delay. However, since HDM will free-stream out of small-scale overdensities, oscillations of the baryon-photon fluid can erase these overdensities. This can lead to a negative transfer function, representing scales where initially overdense regions have become underdense by $z = 199$, and vice versa \citep{Wittenburg_2023}. This is not problematic theoretically, since \autoref{eq:Power Transfer relation} shows that the phase of the transfer function is unimportant for the power spectrum. Logistically though, this process requires us to make a slight modification to the \textsc{music} code to read in the magnitude of the transfer function. This is because the logarithm of the transfer function is used, so the value must be positive.

Once $T(k)$ has been read in, \textsc{music} will generate initial particle positions and velocities by convolving $T(k)$ with random Gaussian white noise \citep{Hahn_2011}. All simulations presented here were generated using the same random seed. Some further parameters provided in the \textsc{music} ``.conf'' file are:
\begin{multline}
    \texttt{boxlength} = 800 \notag \\
    \texttt{zstart} = 199 \notag \\
    \texttt{levelmin} = 8 \notag \\
    \texttt{levelmax} = 10 \notag \\
    \texttt{baryons} = no \notag \\
    \texttt{periodic\_TF} = yes. \notag \\
\end{multline}
This will generate a box with sides of $800/h$ comoving Mpc (cMpc) at a starting redshift of $z_{i} = 199$, which is set by the ``zstart'' parameter. The ``baryons = no'' option sets the simulation to assume all the matter is collisionless, which should be an appropriate approximation on the scales we investigate. ``levelmin'' defines the base coarse grid for the box and also the number of simulation particles, which is $N_{\mathrm{part}} = 2^{3 \, \mathrm{levelmin}}$. Hence these simulations have $256^{3}$ particles and a mass per particle of $2.67 \times 10^{12} h^{-1} \, M_\odot$. The maximum spatial resolution is defined by ``levelmax'' and will be $781/h$~ckpc in this case. The ``periodic\_TF'' option gives the simulation periodic boundary conditions.

The authors of \citet{Wittenburg_2023} made some slight adjustments to the output files of \textsc{music} so that they are compatible with the simulation code used. The modified version of \textsc{music} used here is publicly available.\footnote{\url{https://bitbucket.org/SrikanthTN/bonnpor/src/master/Cosmo_patch_and_setup_and_halofinder/music/}}

\subsection{Simulation code}

The simulations were performed using \textsc{phantom of ramses} \citep*[\textsc{por};][]{Lughausen_2015}, a modified version of the Adaptive Mesh Refinement (AMR) code \textsc{ramses} \citep{Teyssier_2002}. We used the collisionless $N$-body mode, which enables only gravitational interactions and treats all particles as point-like. MOND is integrated into the \textsc{por} code by solving the ordinary Poisson equation twice, once to obtain $\bm{g}_{_{\mathrm{N}}}$ and another time to solve \autoref{eq:div_g_gN}, with a simple algebraic calculation and numerical differentiation in between.

The authors of \citet{Wittenburg_2023} made further modifications to this code so that it can be applied to cosmological simulations, in particular by using supercomoving coordinates \citep{Martel_1998, Teyssier_2002}. Comoving coordinates provide a more natural coordinate system to incorporate the expansion of the universe, but it is also convenient to apply several other scalings inspired by similar modifications made to the \textsc{raymond} code \citep{Candlish_2015, Candlish_2016}. For instance, comoving positions are divided by the comoving box size, the comoving density is divided by the average for the simulation box, and times are in units of the present Hubble time $1/H_0$ \citep[see appendix~A of][]{Wittenburg_2023}. More details on the specifics of these overall modifications are provided in \citet{Wittenburg_2020} and \citet{Wittenburg_2023}, while \citet{Nagesh_2021} provides a user guide for \textsc{por}.

\section{Analysis of simulation outputs}
\label{sec:Analysis}

We use the \textsc{extract\_por} tool to extract the individual particle data from the binary simulation outputs and represent them in ASCII form.\footnote{\url{https://bitbucket.org/SrikanthTN/bonnpor/src/master/extract_por/}} The front bottom left corner of the simulated domain is at the origin, so the $x$, $y$, and $z$ coordinates are all $\geq 0$.

Due to computational and time constraints, locally inferred parameters were measured from a subset of $25^{3}$ particles, which we call ``vantage points'' (VPs). We run analyses at different VPs in parallel, with the computational efficiency further improved by using a basic tree code in which particles are sorted into $100^3$ voxels \citep[this is similar to the approach used in][whose tree code implementation is publicly available]{Banik_2021_backsplash}. The VPs are chosen by splitting the simulation volume into $25^{3}$ voxels and then searching for the particle closest to the centre of each voxel. We require the VP particle to be within $16/h$~Mpc of the voxel centre to limit the overlap between any two VPs, which is another reason not to consider a too dense grid of VPs. Considering results from a simulation particle should provide a fair representation of the simulation as a whole, without overly biasing them to, e.g., overdense regions where more particles are present. Requiring each VP to reside at a particle should also provide a fairer comparison to what a real observer could see -- it would be very difficult for us to measure anything from a region with no galaxies to live in! 

Once we have selected the VPs from which to do our analyses, we then calculate the locally inferred density contrast, bulk flow, apparent Hubble constant and deceleration parameter, and Hubble dipole that an observer would measure from the VP. We detail the methodology for these calculations below in Sections~\ref{sec:Density contrast}, \ref{sec:Bulk flows}, \ref{sec: Hubble constant}, and \ref{sec: Hubble dipole}, respectively. We also apply a group-finding algorithm to identify haloes that are likely gravitationally bound (Section~\ref{sec:Halofinder}).

\subsection{Halo counts}
\label{sec:Halofinder}

A simple way to quantify structure formation is with the Cumulative Halo Mass Function (CHMF), the comoving number density of haloes above some specified mass. This $n \left( M \right)$ has been determined observationally by studies such as \citet{Driver_2022} and \citet{Wang_2022_DESI} using the Galaxy and Mass Assembly \citep[GAMA;][]{Driver_2009} survey and the Dark Energy Spectroscopic Instrument (DESI) Legacy Imaging Survey DR9 \citep{Dey_2019}, respectively.

We choose \citet{Driver_2022} as the observational reference because they probe to higher masses, better complementing our large-scale simulations. Their method fits the observed halo mass function to the `MRP function' given by \citet*{Murray_2018}:
\begin{eqnarray}
    \label{eq:MRP HMF}
    &&\phi \left[\log_{10}(M/M_{\odot})\right] \equiv \frac{dn}{d(\log_{10}(M/M_{\odot}))} \nonumber \\
    &=& \ln(10)\phi_{*}\beta\left(\frac{M}{M_{*}}\right)^{\alpha+1}\exp\left[-\left(\frac{M}{M_{*}}\right)^{\beta}\right].
\end{eqnarray}
We use their `Omega' parameter fit (which enforces the \citealt{Planck_2020} $\Omega_{\mathrm{m}}$ as a prior) and correct to $z=0$ such that $\log_{10} \phi_* = -4.565^{+0.29}_{-0.24} \text{Mpc}^{-3}\text{dex}^{-1}$, $\log_{10}\left( M_{*}/M_{\odot} \right) = 14.505^{+0.11}_{-0.15}$, $\alpha = -1.85^{+0.03}_{-0.03}$, and $\beta = 0.77^{+0.11}_{-0.11}$. The CHMF is then found by integrating \autoref{eq:MRP HMF} in log-space to $+\infty$ such that
\begin{eqnarray}
    \label{eq:HMF integral}
    N \left( >M \right) ~=~ \int_{M}^{+\infty} \phi\left[\log_{10} M^{\prime} \right]\,d(\log_{10} M^{\prime}).
\end{eqnarray}
This observed CHMF is stated to be valid in the range $10^{12.5}M_{\odot}/h < M < 10^{15.3}M_{\odot}/h$.

To identify haloes in the simulations, we use the \textsc{HOP} \citep{Eisenstein_HOP} group-finding algorithm built into the \textsc{yt\_astro\_analysis} extension \citep{Smith_yt} of the \textsc{yt} analysis toolkit \citep{Turk_2011}. This is a density-based group-finding algorithm which links particles to a group using their highest density neighbour. The local density at each particle is calculated by the radial distance to the closest $N_{\mathrm{dens}}$ particles. We use a base value of $N_{\mathrm{dens}} = 65$. Particles are linked into ``chains'' using their highest density neighbour, until a particle is reached that has a higher density than all of its neighbours. A group about this particle is formed from all the chains that connect to it. These groups are then accepted or rejected based on an arbitrary density threshold, which we take to be an overdensity of $200\times$ the critical density $\rho_c$ to match \citet{Driver_2022}. The algorithm therefore outputs a mass of $M_{200}$ and corresponding virial radius of $R_{200}$, which are related by 
\begin{eqnarray}
    \label{eq:Virial mass}
    M_{200} ~\equiv~ \left( 200 \, \rho_c \right) \left( \frac{4\mathrm{\pi} R_{200}^{3}}{3} \right).
\end{eqnarray}
A few other useful quantities are also extracted.

As discussed by \citet{Wittenburg_2023}, algorithms such as \textsc{HOP} are built for simulations using Newtonian gravity, with no equivalent algorithms yet made for MOND gravity. We follow their example and report a Newtonian dynamical mass $M_N$ for the structures identified.
\begin{eqnarray}
    \label{eq:Dynamical mass}
    M_{\mathrm{N}} ~=~ \nu \left( \frac{g_{_{\mathrm{N}}}}{{a_{_0}}} \right) M_{200} \, ,
\end{eqnarray}
where $g_{_{\mathrm{N}}}$ is evaluated at the virial radius $M_{200}$, giving
\begin{eqnarray}
    g_{_{\mathrm{N}}} ~=~ \frac{GM_{200}}{R_{200}^{2}} \, .
\end{eqnarray}
This modification was inspired by a similar technique used by \citet{Angus_2011_satellites}. It allows for a fairer comparison between the MOND simulations and observations. All halo masses presented for the MOND simulations will be these Newtonian dynamical masses, which better correspond to the mass that observers would report in the literature using probes of the internal dynamics or lensing.

This general boost to the gravity caused by MOND likely also means that there are bound structures which are overlooked by the \textsc{HOP} algorithm, which a specialized MOND groupfinder would detect. The number counts reported by \textsc{HOP} may therefore be better interpreted as a lower bound on the true number counts for the MOND simulations. 

Previous work has suggested that simulations with warm DM (WDM) or HDM may produce spurious small-scale structures \citep{Colin_2008, Power_2016}. To curb the effect of any such small-scale issues, we conservatively require all considered haloes to have $\geq 100$ particles.

Once the haloes have been identified, the simulated CHMF can be calculated by dividing the number of haloes above a given mass by the comoving simulation volume. The results are shown in Section~\ref{sec:Halo mass functions}.

\subsection{Density contrast}
\label{sec:Density contrast}

The simplest location-dependent parameter is the average density in some suitably chosen volume centred on each VP. Since we also know the average density $\overline{\rho}$ in our simulation box from the adopted cosmological parameters or analysis of the results, the first locally inferred parameter we measure is the fractional density contrast $\delta$. In order to better replicate the observations taken by \citet*{Keenan_2013} and \citet{Wong_2022}, we only consider particles within $R_{\delta} = 225/h$~Mpc and beyond $r_\delta = 27/h$~Mpc of each VP, ignoring particles outside this distance range. This is to replicate the redshift range of \citet{Wong_2022}. We also mask a randomly oriented region of the sky at each VP to mimic the obscuring effect of the Galactic disc. We follow the procedure laid out by \citetalias{Haslbauer_2020}, which we briefly revisit.

We begin by setting the visible area $A$ of the sky equal to the survey area, which in our case is $A = 37080$~deg\textsuperscript{2} as stated in section~2.5 of \citet*{Keenan_2013}.\footnote{This is very similar to the 37063~deg\textsuperscript{2} stated in table~1 of \citet{Wong_2022}.} The direction of the masked region is randomly chosen for each VP based on a random direction, which is analogous to the north Galactic pole. For this, we sample two random numbers $u_\theta$ and $u_\phi$ that are uniform over the range $0-1$, i.e., $u_{\theta}, u_{\phi} \sim \mathcal{U} \left( 0,1 \right)$. We then calculate the polar angle $\theta$ and azimuthal angle $\phi$ as follows:
\begin{eqnarray}
    \label{eq:random angles disc}
    \cos \theta ~&=&~ 2u_{\theta} - 1 \, , \\
    \phi ~&=&~ \cos \left( 2\mathrm{\pi} u_{\phi} \right).
\end{eqnarray}
The corresponding unit vector $\widehat{\bm{n}}$ representing the Galactic spin axis is then
\begin{eqnarray}
    \label{eq:unit vector disc}
    \widehat{\bm{n}} \equiv \begin{pmatrix}
    \widehat{\bm{n}}_{x} \\
    \widehat{\bm{n}}_{y} \\
    \widehat{\bm{n}}_{z} \end{pmatrix}
    ~=~ \begin{pmatrix}
    \sin \theta \cos \phi \\
    \sin \theta \sin \phi \\
    \cos \theta
    \end{pmatrix}.
\end{eqnarray}

Since the Galactic disc is positioned in the plane perpendicular to $\widehat{\bm{n}}$, observers see galaxies only close to $\pm \widehat{\bm{n}}$. The angular radius of the masked region near the Galactic equator can be calculated by relating the surface area element and solid angle element of a unit sphere. This relation can be expressed as the following integral:
\begin{eqnarray}
    \label{eq:Area - Solid angle relation}
    A ~=~ 2\int_{0}^{2\mathrm{\pi}} \int_{0}^{\frac{\mathrm{\pi}}{2} - \omega_{\delta}}  \sin \theta \, d\theta \, d\phi \, ,
\end{eqnarray}
where $\omega_{\delta}$ is the half-angle width of the sky region obscured by the Galactic disc, or equivalently the minimum observable Galactic latitude. Evaluating this integral and rearranging gives
\begin{eqnarray}
    \label{eq:angle delta}
    \cos \omega_{\delta} ~=~ 1 - \frac{A}{4\mathrm{\pi}} \, ,
\end{eqnarray}
which matches equation~14 of \citetalias{Haslbauer_2020}.

A particle $p$ is accepted into the analysis if its position $\bm{r}_{p}$ relative to the VP is within $\frac{\mathrm{\pi}}{2} - \omega_{\delta}$ of the axis defined by $\pm \widehat{\bm{n}}$, taking the sign that gives the smaller angle. This can be expressed mathematically through the dot product as follows:
\begin{eqnarray}
    \label{eq:delta mask condition}
    \left| \bm{r}_{p} \cdot \widehat{\bm{n}} \right| ~>~ r_p \cos \omega_{\delta} \, ,
\end{eqnarray}
where $r \equiv \left| \bm{r} \right|$ for any vector $\bm{r}$ in what follows, so in this case $r_p$ is the distance of particle $p$ from the observer. The modulus is required on the left hand side because valid positions unobscured by the mock Galactic disc are found both above and below it. Any particle that does not meet this condition is masked by the mock Galactic disc and ignored.

We find the total mass $M$ of all particles within $R_\delta$ and beyond $r_\delta$ that are unobscured by the mock Galactic disc. The volume of the considered region is thus
\begin{eqnarray}
    V ~=~ \frac{2\mathrm{\pi} \left( R_\delta^{3} - r_\delta^3 \right)}{3} \left( 1 + \cos \omega_{\delta} \right).
\end{eqnarray}
The locally inferred density contrast is then
\begin{eqnarray}
    \label{eq:density contrast}
    \delta ~\equiv~ \frac{M}{V\overline{\rho}} - 1.
\end{eqnarray}
We show our simulated density contrasts in Section~\ref{sec:Density contrast results}.

\subsection{Bulk flows}
\label{sec:Bulk flows}

The final particle level analysis we perform is calculating bulk flows so that they can be compared to \citet{Watkins_2023}. We mimic their approach and draw a sphere around each VP, split into equally spaced radial bins of width $25/h$~Mpc. The innermost bin of each sphere extends up to $r_{\mathrm{bf,in}} = 25/h$~Mpc and the outermost bin extends up to $r_{\mathrm{bf,out}} = 400/h$~Mpc, for a total of 16 bins. Measuring out to such large radii also allows us to check the possibility of a bulk flow sourcing the Hubble dipole observed by \citet{Migkas_2021}.

To create an equivalent comparison to \citet{Watkins_2023}, we calculate bulk flows using only the line of sight (LOS) or radial peculiar velocity $v_{r,p}$ of particle $p$ from each VP (\autoref{eq:vr_b}). This requires us to first calculate the unit vector $\widehat{\bm{n}}_p$ of particle $p$ from the VP.
\begin{eqnarray}
    \label{eq:bf unit vector}
    \widehat{\bm{n}}_p ~\equiv~ \frac{\bm{r}_p}{r_p} \, .
\end{eqnarray}
Although $v_{r,p}$ is usually considered a scalar quantity, we will interpret it as a vector by assigning it the LOS direction.
\begin{eqnarray}
    \label{eq:radial velocity bulk flow}
    \bm{v}_{r,p} ~=~ v_{r,p} \widehat{\bm{n}}_{p} \, .
\end{eqnarray}
The bulk flow at radius $R$ can then be calculated as 
\begin{eqnarray}
    \label{eq:bulk flow}
    \bm{v}_{\mathrm{bf}} \left( R \right) ~=~ \frac{3\sum_{p \left( <R \right)} \left( \frac{M_{p} \bm{v}_{r,p}}{r_p^{2}} \right)}{\sum_{p\left( <R \right)} \left(\frac{M_{p}}{r_p^{2}} \right)} \, ,
\end{eqnarray}
where $\sum_{p \left( <R \right)}$ represents a sum over all particles within radius $R$. The factor of 3 here comes from equation~2 of \citet{Nusser_2016}. It ensures that if all particles within the considered volume had the same $\bm{v}$, then the inferred bulk flow would be $\bm{v}_{\mathrm{bf}} = \bm{v}$. The projection of $\bm{v}$ onto the LOS and the projection of the LOS onto the direction of $\bm{v}$ result in an average factor of 1/3 \citep{Mazurenko_2024}. This is cancelled out by introducing an extra factor of 3 into the definition of the bulk flow. Therefore, we can approximately equate $\bm{v}_{\mathrm{bf}}$ with the mass-weighted mean velocity of all the matter within $R$, even though strictly speaking this requires knowledge of 3D velocities and thus cannot be obtained by observers using only LOS peculiar velocities. A major advantage of using bulk flows rather than peculiar velocities is that the latter can only be obtained from the independent observables of redshift and distance if we know $H_0$, which however is highly controversial. But since the bulk flow is an average velocity within a spherical region, it is insensitive to the assumed value of $H_0$. This is because roughly speaking, it measures the dipole angular variation in redshift at fixed distance.

Our simulated bulk flows are shown in Section~\ref{sec:Results bulk flows}, with a comparison to the reported angle between the observed Hubble dipole and local bulk flow found in Section~\ref{sec:Results Hubble dipole}.

\subsection{Hubble constant and deceleration parameter}
\label{sec: Hubble constant}

The background Hubble constant in our simulations is $H_0 = 67.4$~km/s/Mpc and the background deceleration parameter is
\begin{eqnarray}
    q_0 ~\equiv~ -\frac{\ddot{a}a}{\dot{a}^2} ~=~ \frac{\Omega_{\mathrm{m}}}{2} - \Omega_\Lambda ~=~ 0.53 \, .
    \label{eq:q_0}
\end{eqnarray}
However, systematic effects due to large-scale structure, such as outflows from a local supervoid, may offset an observers' measurement of these parameters from their true values (Section~\ref{sec:nuHDM model}). In what follows, we use $\bm{v}$ to denote a peculiar velocity. Following the procedure of \citetalias{Haslbauer_2020}, we construct a quadratic model for $v_r$, the radially outward component of $\bm{v}$ from a particular VP. We calculate results at thousands of different VPs to explore what the universe would look like from different viewpoints.

Since we have a large number of particles, we obtain the mean $v_r$ in shells, yielding a curve of $\overline{v}_r \left( r \right)$, where $r$ is the distance from the observer. We begin by constructing a hollow sphere with inner radius $r_{H,\rm{in}} = 70/h$~Mpc ($z \approx 0.023$) and outer radius $r_{H,\rm{out}} = 400/h$~Mpc ($z \approx 0.14$) around each VP, comparable to the redshift range used in \citealt{Camarena_2020a} ($0.023 < z < 0.15$), with whom we compare. Since our simulations have a box length of $800/h$~Mpc, we cannot further extend the outer radius in this analysis without double-counting particles. We split each hollow sphere into ten equally spaced radial bins of width $33/h$~Mpc. For any given bin $b$, we define its mass-weighted distance from the observer as
\begin{eqnarray}
    \overline{r}_{b} ~\equiv~ \frac{\sum_{p} M_{p} r_p}{\sum_{p} M_{p}} \, ,
    \label{eq:r_b}
\end{eqnarray}
where an overline denotes a mass-weighted average, $p$ is the index of each particle with position vector $\bm{r}_{p}$ relative to the VP, and $\sum_{p}$ is a sum over all particles in bin $b$. Similarly, we find the mass-weighted $v_r$ as follows:
\begin{eqnarray}
    \overline{v}_{b} ~\equiv~ \frac{\sum_{p} M_{p} \overbrace{\left( \bm{r}_{p} \cdot \bm{v}_{p}/r_{p} \right)}^{v_{r,p}}}{\sum_{p} M_{p}} \, ,
    \label{eq:vr_b}
\end{eqnarray}
omitting the $r$ subscript on $\overline{v}$ for clarity and using $\bm{v}_{p}$ for the peculiar velocity vector of particle $p$. Peculiar velocities in the simulation frame are first corrected for the overall barycentre peculiar velocity and therefore correspond to observed velocities in the CMB frame, which is the usual convention because the velocity of the Sun relative to the CMB is known very precisely \citep{Smoot_1977, Kogut_1993}. Observers therefore typically convert velocities in the less fundamental heliocentric frame to the more fundamental CMB frame \citep[see also][]{Wagenveld_2024}.

We now fit the binned outward peculiar velocity $\overline{v}_r$ using a parabolic dependence on $\overline{r}$:
\begin{eqnarray}
    \overline{v}_{b} \left( \overline{r} \right) ~=~ \mathcal{A}_{H} \overline{r}_{b}^{2} + \mathcal{B}_{H} \overline{r}_{b} \, ,
    \label{eq:quadratic hubble}
\end{eqnarray}
where $\mathcal{A}_{H}$ and $\mathcal{B}_{H}$ are free parameters to be fit for each VP, allowing us to gauge the relation with other observer-dependent quantities like the local density contrast (Section~\ref{sec:Density contrast}). Our approach is different from standard polynomial fitting because we force the intercept to be zero. This is because physically the outward velocity at zero distance from any observer must be zero to have a continuous velocity field. To find the best-fitting coefficients $\mathcal{A}_{H}$ and $\mathcal{B}_{H}$, we construct a quadratic loss function $\lambda$ as follows:
\begin{eqnarray}
    \label{eq:loss function hubble}
    \lambda \{ \overline{r}_{b}, \overline{v}_{b} \} ~=~ \sum_{b} \left[\overline{v}_{b} - \left( \mathcal{A}_{H}\overline{r}_{b}^{2} + \mathcal{B}_{H}\overline{r}_{b} \right) \right]^{2} \, ,
\end{eqnarray}
where $\sum_{b}$ is a sum over all bins. The optimal solution for $\mathcal{A}_{H}$ and $\mathcal{B}_{H}$ is found when $\lambda$ is minimized, requiring that $\frac{\partial \lambda}{\partial \mathcal{A}_{H}} = 0$ and $\frac{\partial \lambda}{\partial \mathcal{B}_{H}} = 0$. Calculating these partial derivatives produces the following simultaneous equations:
\begin{eqnarray*}
    \label{eq:quadratic simulataneous equation}
    \sum_{b} \overline{r}_{b} \overline{v}_{b} &=& \mathcal{A}_{H} \sum_{b} \overline{r}_{b}^{3} + \mathcal{B}_{H} \sum_{b} \overline{r}_{b}^{2} \\
    \sum_{b} \overline{r}_{b}^{2} \overline{v}_{b} &=& \mathcal{A}_{H} \sum_{b} \overline{r}_{b}^{4} + \mathcal{B}_{H} \sum_{b} \overline{r}_{b}^{3}.
\end{eqnarray*}
We use Gaussian elimination to solve these simultaneous equations for $\mathcal{A}_{H}$ and $\mathcal{B}_{H}$.

We next use these coefficients to estimate $\mathcal{H}_0$ and $\mathcal{Q}_0$, the apparent Hubble and deceleration parameters deduced assuming only cosmological expansion contributes to the redshift. In reality, the wavelength of a photon gets stretched both due to cosmological expansion and outward peculiar velocity $v_r$, so the redshift $z$ satisfies
\begin{eqnarray}
    \left( 1 + z \right) ~=~ \left( 1 + z_c \right) \left( 1 + v_r/c \right),
\end{eqnarray}
where $z_c \equiv a^{-1} - 1$ is the purely cosmological contribution to $z$. We can invert this relation to find the apparent cosmic scale factor $a_{\mathrm{app}}$ that would be found if redshifts were assumed to arise solely from cosmic expansion:
\begin{eqnarray} 
    a_{\mathrm{app}} ~=~ \frac{1}{1 + z} = \frac{a}{1 + v_r/c} \, .
\end{eqnarray}
Differentiating with respect to time gives
\begin{eqnarray}
    \dot{a}_{\mathrm{app}} &=& \frac{\dot{a}}{1 + v_r/c} - \frac{a}{c \left( 1 + v_r/c \right)^2} \dot{v_r} \\
    &=& \frac{\dot{a}}{1 + v_r/c} + \frac{a}{\left( 1 + v_r/c \right)^2} v_r' \, ,
    \label{eq:a_app_dot}
\end{eqnarray}
where we used the relation $dr = -c \, dt$ for our past lightcone, approximating that $v_r$ has changed little in the light crossing time of any local inhomogeneity. At $a = 1$ and $v_r = 0$, we get that
\begin{eqnarray}
    \mathcal{H}_0 \equiv \dot{a}_{\mathrm{app}} ~=~ \dot{a} + v_r' \, = \dot{a} + \mathcal{B}_{H} \, ,
\end{eqnarray}
substituting in our parabolic binned relation between $v_r$ and $r$ (\autoref{eq:quadratic hubble}). We get the expected result that the linear coefficient $\mathcal{B}_{H}$ acts as an extra contribution to the apparent Hubble constant.

Differentiating \autoref{eq:a_app_dot} with respect to time gives
\begin{eqnarray}
    \ddot{a}_{\mathrm{app}} &=& \frac{\ddot{a}}{1 + v_r/c} - \frac{\dot{a}}{c \left( 1 + v_r/c \right)^2} \dot{v} + \frac{\dot{a}}{\left( 1 + v_r/c \right)^2} v_r' \nonumber \\
    &~&+ \frac{a}{\left( 1 + v_r/c \right)^2} \dot{v}_r' - \frac{2a}{c \left( 1 + v_r/c \right)^3} v_r' \dot{v_r} \\
    &=& \frac{\ddot{a}}{1 + v_r/c} + \frac{2\dot{a}}{\left( 1 + v_r/c \right)^2} v_r' - \frac{ac}{\left( 1 + v_r/c \right)^2} v_r'' \nonumber \\
    &~&+ \frac{2a}{\left( 1 + v_r/c \right)^3} v_r'^2 \, .
\end{eqnarray}
At $a = 1$ and $v_r = 0$, this reduces to
\begin{eqnarray}
    \ddot{a}_{\mathrm{app}} ~&=&~ \ddot{a} + 2 \dot{a} v_r' - cv_r'' + 2 v_r'^2 \, \\
    &=&~ \ddot{a} + 2 v_r' \left( \dot{a} + v_r' \right) - cv_r'' \\
    &=&~ \ddot{a} + 2 \left( \mathcal{B}_H \mathcal{H}_0 - \mathcal{A}_H c \right).
\end{eqnarray}
We then get the background $\ddot{a}$ from \autoref{eq:q_0} and use an analogous relation to obtain the apparent deceleration parameter:
\begin{eqnarray}
    \mathcal{Q}_0 ~\equiv~ -\frac{\ddot{a}_{\mathrm{app}}}{\mathcal{H}_0^2} \, .
\end{eqnarray}
We present the simulated apparent Hubble constant and deceleration parameter in Section~\ref{sec:results Hubble tension}.


\subsection{Hubble dipole}
\label{sec: Hubble dipole}

Some studies have reported a dipole-like angular variation in the apparent Hubble constant across the sky \citep{Migkas_2018, Kalbouneh_2023, Hu_2024a, Hu_2024b, Sah_2025}. We compare our results with the study of \citet{Migkas_2021}, who used galaxy cluster scaling relations to obtain distances to galaxy clusters, albeit without any absolute calibration. This prevents their study from directly addressing the Hubble tension, but it does allow for an interesting cosmological test using the percentage variation in $\mathcal{H}_0$ across the sky. \citet{Migkas_2021} report a Hubble dipole of $9.0 \pm 1.7\%$ on scales of about $350/h$~Mpc, which is of a similar magnitude to the Hubble tension. They suggest that this signal could be sourced by peculiar velocities, rather than being a physical difference in expansion rate across the sky. We test this hypothesis by quantifying the observed dipole induced by peculiar velocities around each VP.

We begin by obtaining the position $\bm{r}_p$ relative to the VP and the radially outward peculiar velocity $v_{r,p}$ (\autoref{eq:vr_b}) for particle $p$. Since the cluster scaling relations of \citet{Migkas_2021} lack any absolute calibration, we need to work out a dipole in percentage terms rather than in km/s/Mpc. This means we need to know the monopole term, which neglecting the Hubble expansion is given by
\begin{eqnarray}
    \label{eq:Monopole Hubble}
    \mathcal{M} \left( R \right) ~=~ \frac{\sum_{p \left( <R \right)} r_p v_{r,p}}{\sum_{p \left( <R \right)} r_p^{2}},
\end{eqnarray}
where $\sum_{p(<R)}$ is the sum over all particles within distance $R$ of the VP. In this case, we set $R = 350/h$~Mpc as this roughly corresponds to the typical distance of the galaxy clusters analysed by \citet{Migkas_2021}.

The dipole $\mathcal{D}$ in the Hubble parameter has the following components:
\begin{eqnarray}
    \label{eq:dipole xyz Hubble}
    \mathcal{D}_{x} \left( R \right) &=& \frac{\sum_{p \left( <R \right)} x_{p}v_{r,p}}{\sum_{p \left( <R \right)} x_{p}^{2}} \, , \\
    \mathcal{D}_{y} \left( R \right) &=& \frac{\sum_{p \left( <R \right)} y_{p}v_{r,p}}{\sum_{p \left( <R \right)} y_{p}^{2}} \, , \\
    \mathcal{D}_{z} \left( R \right) &=& \frac{\sum_{p \left( <R \right)} z_{p}v_{r,p}}{\sum_{p \left( <R \right)} z_{p}^{2}} \, ,
\end{eqnarray}
where $x_{p}$, $y_{p}$, and $z_{p}$ are, respectively, the $x$, $y$, and $z$ components of $\bm{r}_p$. The total inferred Hubble dipole is then simply
\begin{eqnarray}
    \label{eq:dipole tot Hubble}
    \mathcal{D} \left( R \right) ~=~ \sqrt{\mathcal{D}_{x} \left( R \right)^{2} + \mathcal{D}_{y} \left( R \right)^{2} + \mathcal{D}_{z} \left( R \right)^{2}} \, ,
\end{eqnarray}
with the percentage dipole being
\begin{eqnarray}
    \label{eq:dipole percent Hubble}
    \mathcal{D}_{\%} \left( R \right) ~=~ \frac{100 \, \mathcal{D} \left( R \right)}{H_0 + \mathcal{M} \left( R \right)} \, .
\end{eqnarray}
Note that since $\mathcal{D}_{\%}$ is the percentage variation of the apparent Hubble constant, the denominator must include the background cosmological expansion and not just the monopole term of the peculiar velocity field (\autoref{eq:Monopole Hubble}).

Our simulated Hubble dipoles can be found in Section~\ref{sec:Results Hubble dipole}, along with a comparison to the reported angle between the observed dipole and bulk flow directions.

\section{Results}
\label{sec:Results}

\begin{figure*}
    \centering
    \includegraphics[width=\textwidth]{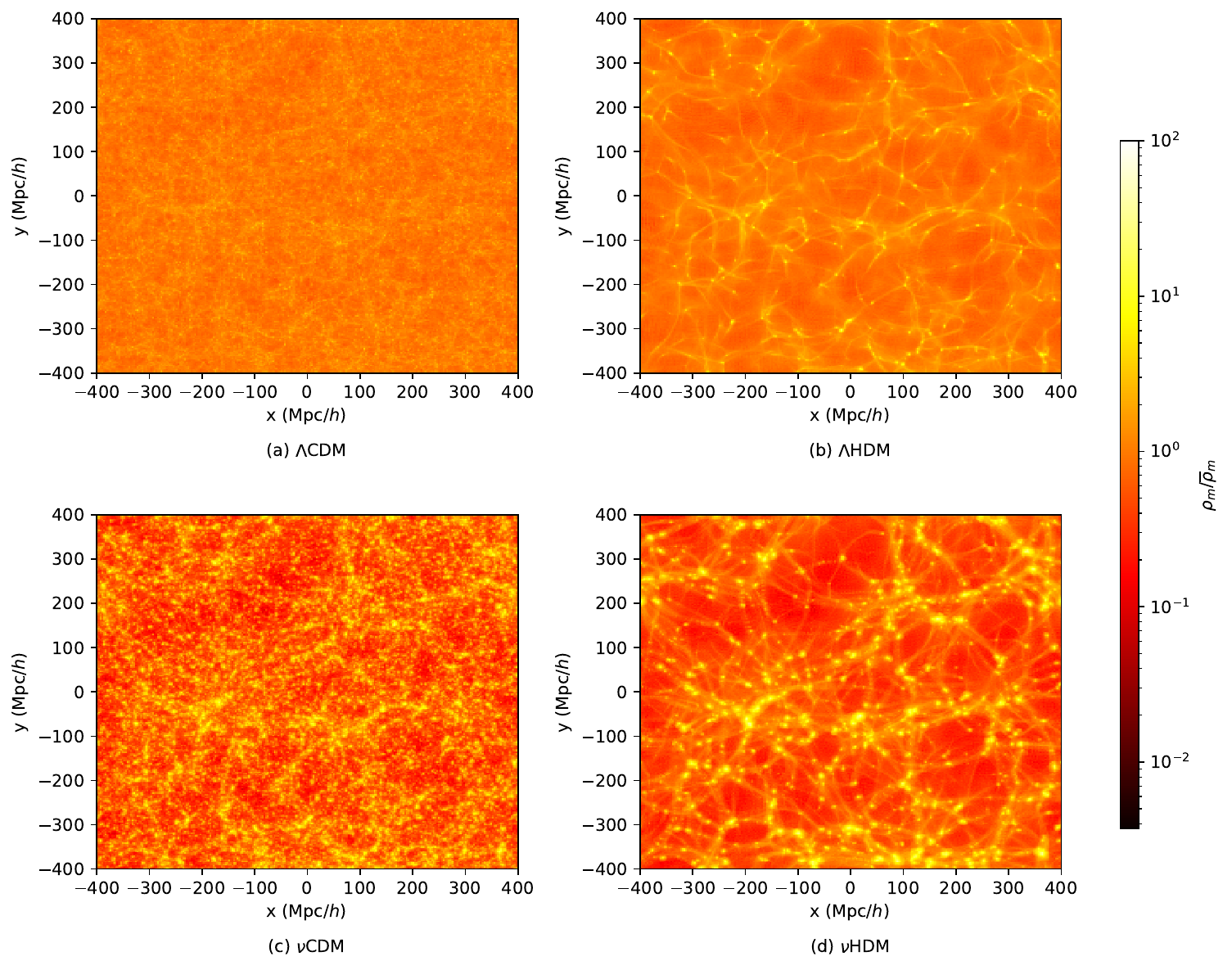}
    \caption{The density projected along the $z$ direction through the whole simulation box at $z = 0$ for $\Lambda$CDM (top left), $\Lambda$HDM (top right), $\nu$CDM (bottom left), and $\nu$HDM (bottom right). The colour of each pixel shows the average 3D density in the pixel in units of the cosmic mean, as indicated using the logarithmic colour bar. The colour scale is identical for each plot.}
    \label{fig:Density maps}
\end{figure*}

We begin by showing final ($z = 0$) snapshots of our four cosmological simulations (\autoref{tab:Model_summary}) in \autoref{fig:Density maps}. The top row shows results using GR, while the bottom row shows the mixed MOND + DM counterpart. The left and right columns show results when using CDM or HDM, respectively, for the DM component. We will continue this formatting for figures going forward to help visualize which changes are due to the gravity law and which are due to the initial power spectrum on small scales.

The $\Lambda$CDM simulation in the top left panel provides a baseline for comparison with the other models. Much more structure is apparent on small scales in the CDM models, presumably because the HDM models have much less initial power on small scales \citep[see figure~1 of][]{Wittenburg_2023}. The gravity law also has a significant impact, with much more structure apparent in the MOND models compared to those which assume GR.

Although $\nu$CDM is unrealistic, it shows a highly structured universe resulting from the combination of both MOND and CDM. At the opposite extreme, the $\Lambda$HDM model shows only modest signs of structure, and even then only on large scales, demonstrating the importance of CDM to structure formation if assuming GR. The $\nu$HDM model represents a middle ground, but it still looks drastically different to $\Lambda$CDM. As discussed in Section~\ref{sec:nuHDM model} and \citet{Wittenburg_2023}, the lack of CDM delays the formation of galaxies. But on large scales, we expect the main difference to be due to the gravity law. Given the very serious problems faced by $\Lambda$CDM on large scales such as the Hubble and bulk flow tensions and the KBC supervoid, we explore the behaviour on large scales further.


\subsection{Peculiar velocities}
\label{sec:pec vel}

\begin{figure*}
    \centering
    \includegraphics[width=\textwidth]{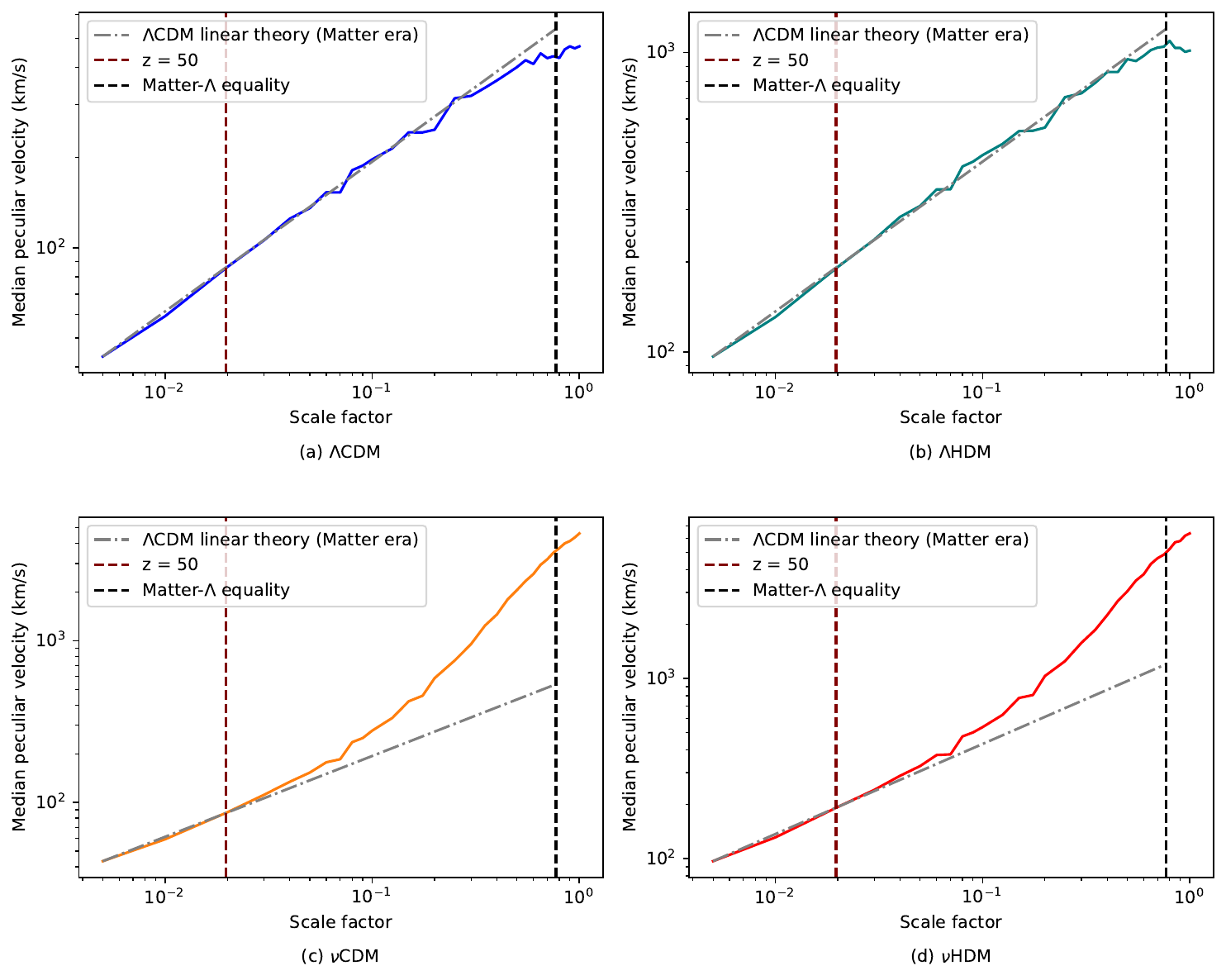}
    \caption{Evolution of the median particle peculiar velocity $v$ with the cosmic scale factor $a$ for $\Lambda$CDM (top left), $\Lambda$HDM (top right), $\nu$CDM (bottom left), and $\nu$HDM (bottom right). Results are shown on logarithmic axes. The grey dash-dotted line shows the predicted $v \propto \sqrt{a}$ evolution from linear Newtonian theory in the absence of dark energy, the maroon dashed line shows the $z = 50$ epoch at which overdensities are predicted to enter the MOND regime \citepalias{Haslbauer_2020}, and the black dashed line shows the epoch of matter--dark energy equality.}
    \label{fig:med pec vel}
\end{figure*}

Structure growth is also evident in the peculiar velocities $v$, which are order several hundred km/s today despite being just a few km/s at recombination, as evidenced by the $10^{-5}$ level temperature fluctuations in the CMB implying peculiar velocities of order $10^{-5} \, c$. We therefore explore the evolution of the peculiar velocity field. For this, we show the median particle peculiar velocity as a function of $a$ (\autoref{fig:med pec vel}). Linear perturbation theory in GR tells us that the density contrasts grow as $\delta \propto a$ and $v \propto \sqrt{a}$ in the matter-dominated regime, which applies to the majority of cosmic history. Indeed, the top panels of \autoref{fig:med pec vel} show that this scaling is accurate until the time when dark energy starts to become important, which we indicate with a dashed vertical line towards the right of each panel. Dark energy slows down the growth of structure due to enhanced Hubble drag.

The MOND models also follow the $v \propto \sqrt{a}$ scaling in the first part of our simulation. This is because the accelerations typically exceed $a_{_0}$ when $z \ga 50$ \citepalias[see section~3.1.3 of][]{Haslbauer_2020}. At later times, the typical acceleration falls below $a_{_0}$, largely putting density perturbations into the MOND regime. As a result, the growth of structure in each MOND model becomes much more rapid than in its GR counterpart, which we show in the panel immediately above. Even so, the MOND models do show a modest deceleration in structure growth at late times due to dark energy, similarly to the models using GR.

\begin{figure*}
    \centering
    \includegraphics[width=\textwidth]{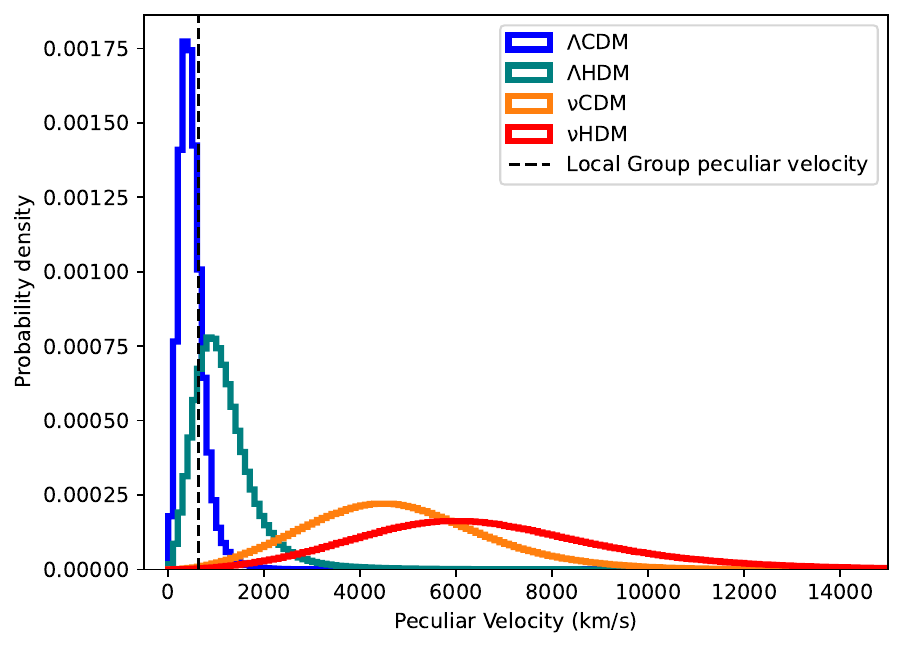}
    \caption{Histograms showing the distribution of particle peculiar velocities in each simulation at $z = 0$. From the leftmost peak to the rightmost peak, these are $\Lambda$CDM, $\Lambda$HDM, $\nu$CDM, and $\nu$HDM. The Local Group peculiar velocity of 627~km/s is shown for reference using a solid vertical line \citep{Kogut_1993}. The typical peculiar velocities in the MOND simulations are thousands of km/s.}
    \label{fig:hists pec vel}
\end{figure*}

To characterise the peculiar velocity distributions in more detail, we show them at $z = 0$ in \autoref{fig:hists pec vel}. We add a vertical black line at the observed Local Group (LG) peculiar velocity of $v_{\mathrm{LG}} = 627$~km/s \citep{Kogut_1993}. $\Lambda$CDM and $\Lambda$HDM (two leftmost peaks) produce present peculiar velocities of a few hundred km/s and can therefore easily account for this observation, but $\nu$CDM and $\nu$HDM (two rightmost peaks) instead yield a few thousand km/s \citep[see also figures~14--16 of][]{Candlish_2016}. The peculiar velocities in the MOND models are so large that it is quite unlikely to find ourselves in an environment with as low a peculiar velocity as the LG. In particular, only 30556 (9931) particles have $v \leq v_{\mathrm{LG}}$ in the $\nu$CDM ($\nu$HDM) simulation, which given the total number of particles would make the LG a $2.9\sigma$ ($3.2\sigma$) outlier. While not conclusive, this is the first indication that $\nu$HDM overcorrects the issues faced by $\Lambda$CDM on large scales. In other words, a more accurate description of nature would not enhance the growth of structure as much as $\nu$HDM, though it should do so to a limited extent.

\subsection{Cumulative Halo Mass Functions}
\label{sec:Halo mass functions}

\begin{figure*}
    \centering
    \includegraphics[width=\textwidth]{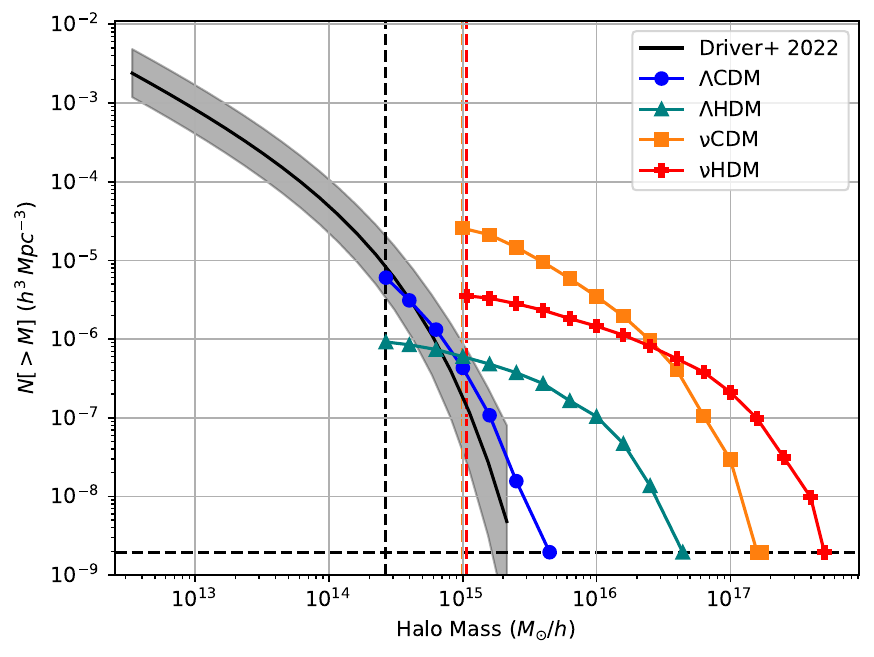}
    \caption{The solid lines with markers show the cumulative halo mass function (CHMF) of each simulation at $z=0$, as indicated in the legend. The first and last markers represent the lightest and heaviest halo mass in each simulation, while the other markers denote intervals of 0.2 dex in mass. The black solid line shows the CHMF observed by \citet{Driver_2022} extrapolated to $z=0$, with the shaded region representing the $1\sigma$ confidence interval. The dashed vertical lines represent the minimum halo mass in each simulation, with the MOND simulations having a higher minimum mass due to the use of Newtonian dynamical masses (\autoref{eq:Dynamical mass}). Poisson uncertainties on simulation results are too small to be visible here. $\Lambda$CDM fits the high mass end of the observations, while the $\Lambda$HDM and MOND simulations appear to overproduce large structures.}
    \label{fig:CHMF}
\end{figure*}

Our simulations form a large number of gravitationally bound objects by $z = 0$ \citep[as also reported by][]{Wittenburg_2023}. We identify these structures using a halo finder (Section~\ref{sec:Halofinder}). Due to the larger size of our simulations and their collisionless nature, these structures correspond to galaxy clusters rather than individual galaxies. We use \autoref{fig:CHMF} to show the CHMF, the comoving number density of haloes above some specified mass. The shape of the simulated CHMF is mostly determined by the assumed nature of the DM component, with CDM models giving a more steeply declining CHMF (compare the blue and orange curves with the red and green curves). This is to be expected because CDM is able to cluster much more efficiently on small scales than HDM. The gravity law mostly affects the normalization, with the enhanced gravity law in MOND shifting the CHMF upwards without much altering the shape (compare the blue and orange curves, or the red and green curves).


The solid black points on \autoref{fig:CHMF} show the observed CHMF \citep{Driver_2022}. This presents a steep decline, which we have seen is characteristic of CDM rather than HDM. The $\Lambda$CDM model appears to fit the most massive end of the observations, where a comparison is possible (our simulations lack the resolution at lower mass). None of the other considered models are as successful in this regard. While it is to be expected that HDM cannot account for the observations in the context of GR gravity, both MOND models are also in severe disagreement with the observations. We confirm with newer observational data the results of earlier simulations by \citet{Angus_2011_clusters} and \citet{Angus_2013}, who also found that $\nu$HDM over-predicts the number of massive halos. We iterate upon their results with our $\nu$CDM and $\Lambda$HDM simulations to better understand the cause of these issues. In particular, we expect that $\nu$CDM and $\nu$HDM bracket the range of possibilities of similar hybrid MOND + DM models, since DM particles with an intermediate mass (WDM) would presumably give a result intermediate between the two. This highlights a likely problem with such hybrid models: the observed steeply declining CHMF requires significant power on small scales. This is not apparent in the CMB, but we also cannot assign it to the DM component because the DM particles should have a rest energy $\la 300$~eV to avoid destroying the MOND fits to galaxy rotation curves \citep{Angus_2010_minimum_neutrino_mass}. Free-streaming effects would then severely limit the power on small scales. One can argue that the power on these scales subsequently grew quite rapidly with MOND gravity. However, this would also cause the power on large scales to grow rapidly, leading to a significant overproduction of the most massive galaxy clusters. There must inevitably be suppression of power on small scales due to DM free-streaming, but the observations imply that this should occur only on very small scales, below the range of cluster masses probed by \citet{Driver_2022}. This is a strong argument in favour of CDM particles regardless of the gravity law.

\subsection{Local density contrasts}
\label{sec:Density contrast results}

An important constraint on any cosmological model is its ability to reproduce the KBC void, which has an apparent underdensity of $46 \pm 6\%$ out to 300~Mpc \citep*{Keenan_2013}. Because of redshift space distortion (RSD) due to outflow from the detected void inflating the redshifts, this corresponds to an $\approx 20\%$ underdensity \citepalias{Haslbauer_2020}. Those authors included an allowance for RSD and compared the observations of \citet*{Keenan_2013} in redshift space to the Millennium~XXL (MXXL) simulation of $\Lambda$CDM \citep{Angulo_2012}. This comparison shows that $\Lambda$CDM fails at $6\sigma$ confidence. The results of \citet*{Keenan_2013} extend out to $800/h_{70}$~Mpc, where $h_{70} \equiv H_0$ in units of 70~km/s/Mpc. This corresponds to $cz = 56000$~km/s or $z = 0.19$, which is sufficient to see the comoving galaxy luminosity density reach a flat level beyond the KBC void (see their figure~11). This extended flat level allows for accurate normalization of the results within the KBC void. However, as discussed in the introduction to \citet{Banik_2025_BAO}, figure~1 of \citet{Wong_2022} shows results out to only $z = 0.08$, by which point the luminosity density is still rising. $z = 0.08$ corresponds to 343/$h_{70}$~Mpc in figure~11 of \citet*{Keenan_2013}, which shows that the luminosity density stops rising at this point, making the density here a good estimate of the cosmic average density $\rho_0$. However, \citet{Wong_2022} assume that the density here is $1.5 \, \rho_0$. Thus, their reported underdensity of the Local Hole ($20 \pm 2\%$) is a significant underestimate. Their published density within the KBC void of $0.8 \, \rho_0$ should actually be rescaled to about $0.8/1.5 = 0.53 \, \rho_0$, which corresponds to a 47\% underdensity -- almost exactly the same as reported by \citet*{Keenan_2013}. This means there is no disagreement between the results of \citet*{Keenan_2013} and the less deep results of \citet{Wong_2022}.

\begin{figure*}
    \centering
    \includegraphics[width=\textwidth]{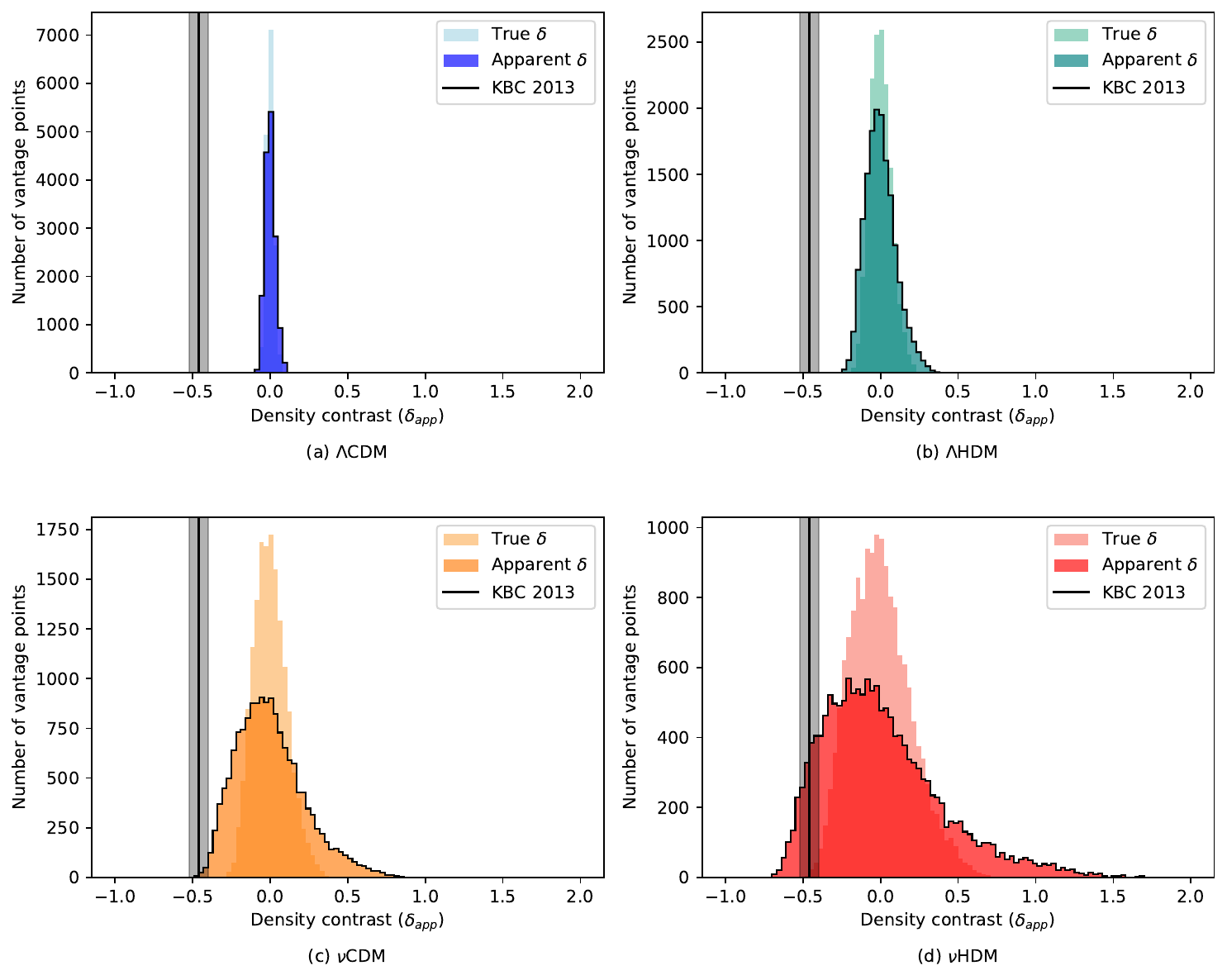}
    \caption{Histograms showing the local density contrast at $z = 0$ within $225/h$~Mpc of each VP for our simulation of $\Lambda$CDM (top left), $\Lambda$HDM (top right), $\nu$CDM (bottom left), and $\nu$HDM (bottom right). Each panel contains two histograms, with the lighter histogram representing the true density contrast and the darker histogram with a black edge representing the apparent (RSD corrected) density contrast an observer would measure (Equation~\ref{eq:redshift_distortion}). The apparent density contrast measured by \citet{Keenan_2013} is shown as a solid black vertical line, with uncertainty indicated using the grey shaded band. The same bins are used in all panels to make comparison easier.}
    \label{fig:hists delta}
\end{figure*}

Both studies report results in redshift space, thereby not accounting for RSD. It is difficult for observers to correct for RSD and find the local density contrast more accurately until redshift-independent distances become available out to $\approx 500$~Mpc. We can however quantify the RSD effect in our simulations using an approach similar to section~3.3.2 of \citetalias{Haslbauer_2020}. For this, we construct a parabolic relation between $cz$ and distance $r$ at the background level, neglecting peculiar velocities. We use this to find $cz$ at $r_\delta$ and $R_\delta$, reasoning that observers must have counted galaxies within this redshift range because they do not correct for RSD, implicitly assuming a homogeneous expansion. Once we have found the linear and quadratic terms at the background level, we add these on to the contributions from peculiar velocities (\autoref{eq:quadratic hubble}). This effectively tells us the relation between recession velocity and distance (though it is not accurate to consider redshifts from cosmological expansion as creating a velocity). We next use this relation to obtain $r_\delta'$ and $R_\delta'$, the inner and outer limit, respectively, to the distance range the observational galaxy number count actually corresponds to for the given VP. The basic idea is that if a local underdensity inflates the local $cz'$ by 10\%, then observers assuming the background redshift-distance relation would find $1.1^3\times$ fewer galaxies than they might expect based on their assumed volume (neglecting the quadratic term). In general, the ratio between the expected and actual volumes is
\begin{eqnarray}
    f_{\mathrm{model}} ~=~ \frac{R_\delta^3 - r_\delta^3}{R_\delta'^3 - r_\delta'^3} \, .
\end{eqnarray}
We can then apply the RSD correction as follows:
\begin{eqnarray}
    \left(1 - \delta_{\mathrm{app}} \right) &=& \left(1 - \delta \right) f_{\mathrm{model}} \, , 
    \label{eq:redshift_distortion}
\end{eqnarray}
where $\delta$ is the actual underdensity in the simulation and $\delta_{\mathrm{app}}$ is the apparent underdensity in redshift space accounting for RSD.

The results obtained with and without correcting the density contrasts in this way are shown in \autoref{fig:hists delta}. Correcting for RSD allows a more direct comparison with the $\delta_{\mathrm{app}} = 46 \pm 6\%$ underdensity reported by \citet{Keenan_2013}, which we show in all panels as a solid black line with grey shaded uncertainty band. The simulated distribution is much narrower in $\Lambda$CDM, but we use the same bins in $\delta_{\mathrm{app}}$ in all panels for ease of comparison. It is clear that $\Lambda$CDM cannot explain the observed $\delta_{\mathrm{app}}$, though interestingly $\Lambda$HDM comes fairly close despite using Newtonian gravity. We attribute this to the model having more power on large scales to compensate for the lack of power on small scales given the lack of CDM. A similar broadening is also apparent in the MOND models when going from $\nu$CDM to $\nu$HDM. This broadening allows $\nu$HDM to readily explain the observed $\delta_{\mathrm{app}}$, while $\nu$CDM can only marginally do so. The ability of only the models based on MOND gravity to match the observed $\delta_{\mathrm{app}}$ provides a strong argument for enhanced structure formation on $\ga 100$~Mpc scales \citepalias[see also][]{Haslbauer_2020}.

\subsection{Bulk flows}
\label{sec:Results bulk flows}

\begin{figure*}
    \centering
    \includegraphics[width=\textwidth]{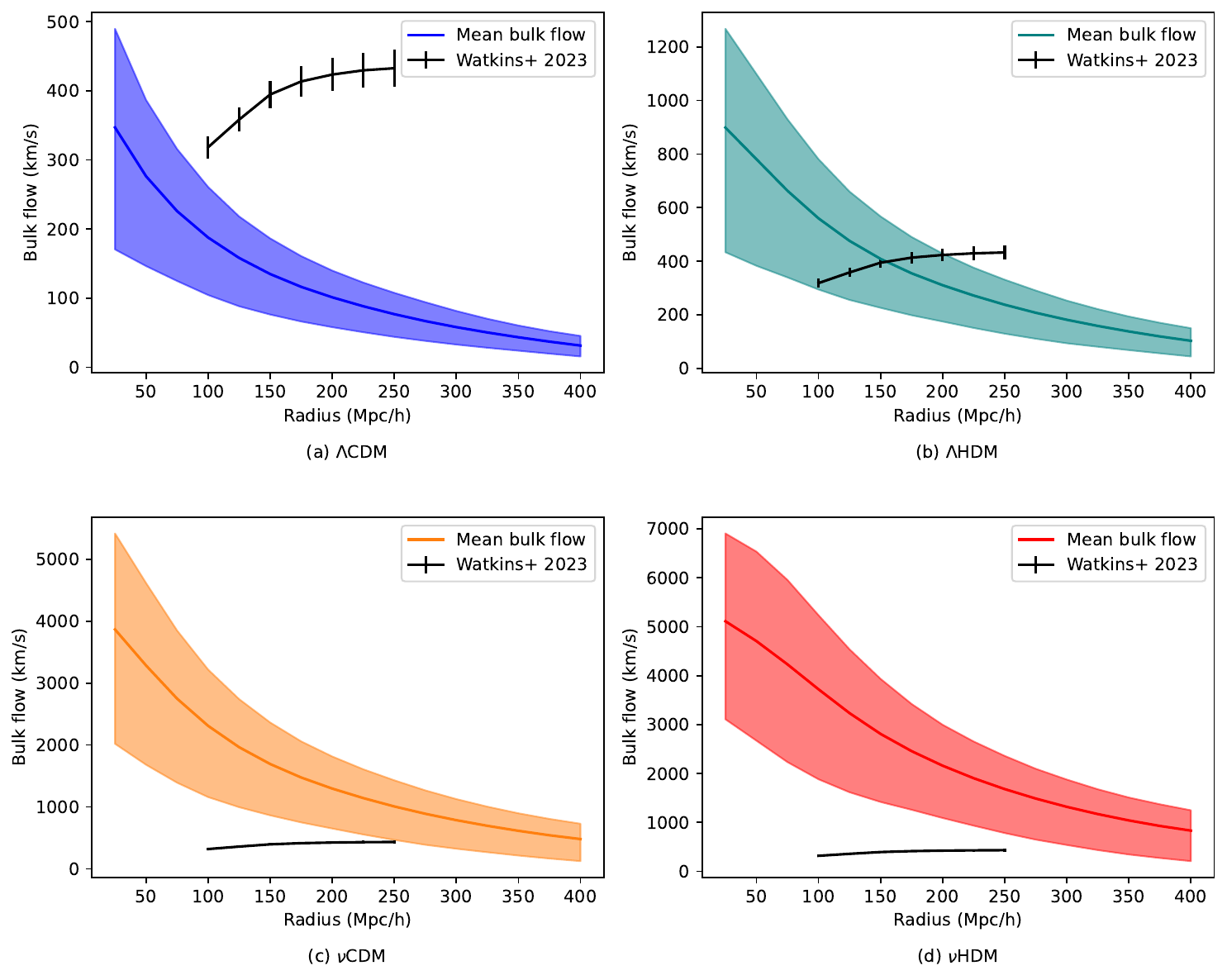}
    \caption{Mean bulk flows for $\Lambda$CDM (top left), $\Lambda$HDM (top right), $\nu$CDM (bottom left), and $\nu$HDM (bottom right). The shaded regions represent the standard deviation from the mean, while the black line with error bars shows the observed local bulk flow \citep{Watkins_2023}. All simulations show a decreasing trend for bulk flow with radius. $\Lambda$CDM predicts much smaller bulk flows than observed at large radii, while the MOND simulations seem to overpredict the bulk flows, especially at low radii. Bulk flows in $\Lambda$HDM seem to have speeds comparable to the observations.}
    \label{fig:Mean bulk flow}
\end{figure*}

Having explored the local density field, we now turn to the local velocity field as traced by the bulk flow curve in each simulation (Section~\ref{sec:Bulk flows}). At each scale, we show the distribution of values across different VPs using the mean and standard deviation (\autoref{fig:Mean bulk flow}). The observed bulk flows from \citet{Watkins_2023} lie well above expectations in $\Lambda$CDM, as quantified in their figure~8. The observed bulk flow is most in line with the $\Lambda$HDM model, with $\nu$CDM appearing to predict too large bulk flows. This is even more so for $\nu$HDM, which predicts that a typical observer should see a bulk flow about $10\times$ larger than observed.

As we go to larger scales, the bulk flow becomes smaller in all panels of \autoref{fig:Mean bulk flow}. This is to be expected because the universe becomes more homogeneous on larger scales in all our models. The expected declining trend is however not observed \citep{Watkins_2023}. Our simulations suggest that extending the observations out to larger radii will show a declining bulk flow \citep[as also suggested by the cosmic dipole in radio source counts above some flux threshold;][]{Wagenveld_2024}. Since the results shown in \autoref{fig:Mean bulk flow} are across all VPs, it may be possible to find some VPs where the bulk flow curve is locally rising, an issue we turn to next.

\begin{table}
    \centering
    \begin{tabular}{cccc}
        \hline
        & \multicolumn{3}{c}{Percentage of VPs which fit observed bulk flow curve} \\
        Simulation & $1\sigma$ & $3\sigma$ & $5\sigma$ \\ \hline
        $\Lambda$CDM & $<0.01\%$ & $<0.01\%$ & $<0.01\%$\\
        $\Lambda$HDM & $0.13\%$ & $0.99\%$ & $2.45\%$ \\
        $\nu$CDM & $<0.01\%$ & $<0.01\%$ & $0.03\%$ \\
        $\nu$HDM & $<0.01\%$ & $<0.01\%$ & $<0.01\%$ \\ \hline
    \end{tabular}
    \caption{Percentage of bulk flow curves in each simulation with a better than $1\sigma$, $3\sigma$, or $5\sigma$ fit to the observed bulk flow curve \citep{Watkins_2023}. Those marked $<0.01\%$ had no matching VPs in our simulations, indicating a fraction $<25^{-3}$.}
    \label{tab:Bulk flow goodness fit}
\end{table}

We consider the bulk flow curve at each VP out to $200/h$~Mpc, beyond which the observational results are less reliable \citep{Watkins_2023}. In particular, the reduced number density of observed galaxies beyond this distance can lead to artificial flattening of the bulk flow curve. We then consider the $\chi^2$ of the simulated bulk flow curve with respect to the observations. We convert the likelihood of a higher $\chi^2$ into a Gaussian equivalent tension for a single variable. In this way, we quantify the proportion of VPs in each simulation for which the observed bulk flow curve is fit to better than $1\sigma$, $3\sigma$, and $5\sigma$. The results are shown in \autoref{tab:Bulk flow goodness fit}. It is clear that rising bulk flow curves similar to that observed can sometimes arise in a simulation that becomes more homogeneous on large scales. The most successful simulation in this respect is $\Lambda$HDM, with $\Lambda$CDM producing bulk flows that are too small and the MOND models producing bulk flows that are too large. This problem is particularly acute for $\nu$HDM. Our results suggest that the \citet{Watkins_2023} rising bulk flow curve is realistic given the initial conditions imprinted by the CMB, but only if structure formation on the relevant $\ga 100$~Mpc scales is enhanced somewhat compared to $\Lambda$CDM. This enhancement should be far smaller than in $\nu$HDM, which cannot be considered a realistic description of the local Universe.

\subsection{Hubble tension and deceleration parameter}
\label{sec:results Hubble tension}

\begin{figure*}
    \centering
    \includegraphics[width=\textwidth]{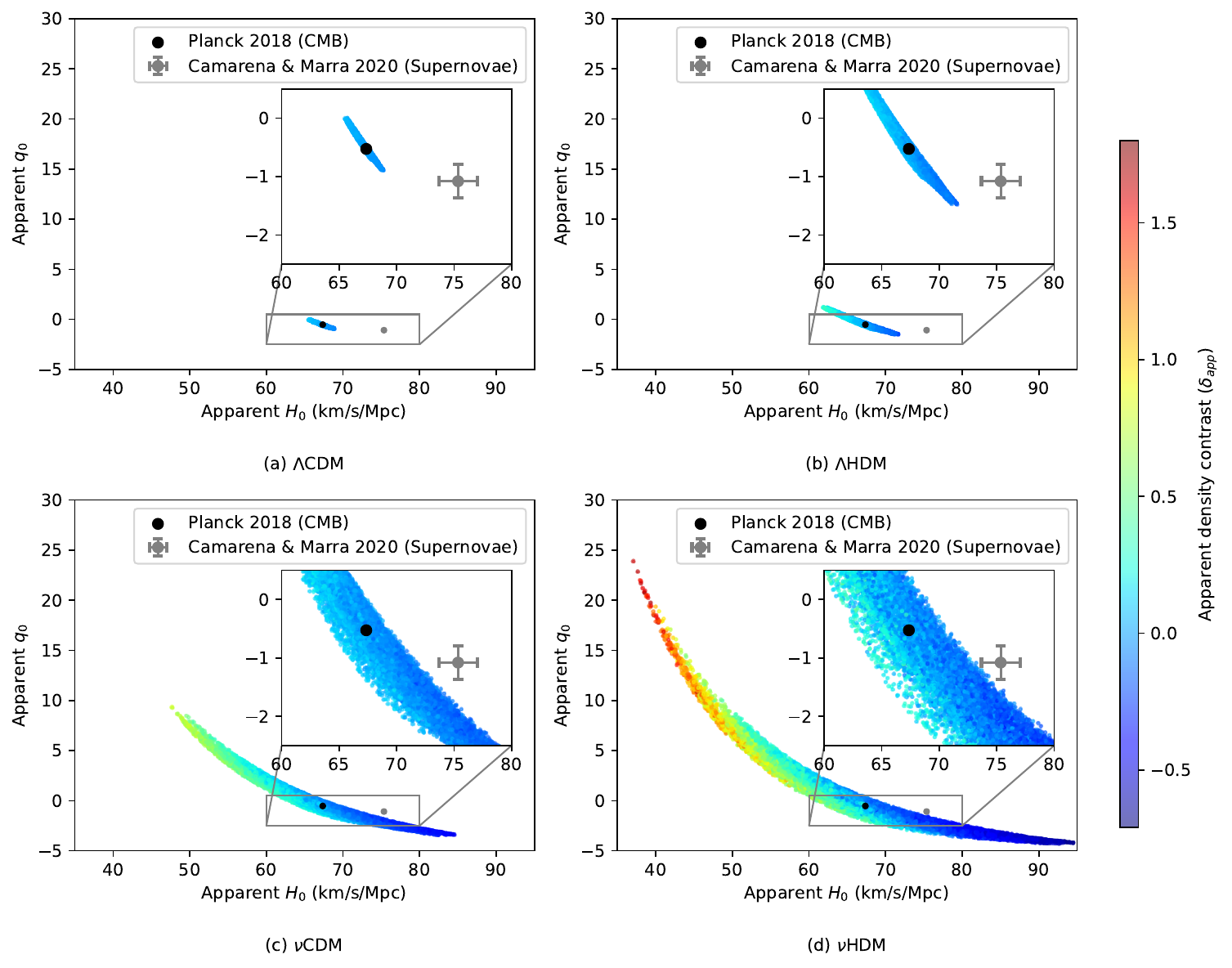}
    \caption{Scatter plots of $q_{0,\rm app}$ vs $H_{0,\rm app}$ for each VP in our simulation of $\Lambda$CDM (top left), $\Lambda$HDM (top right), $\nu$CDM (bottom left), and $\nu$HDM (bottom right). Markers are coloured according to the apparent density contrast about the VP. The values measured by \citet{Camarena_2020a} and \citet{Planck_2020} are shown for reference, with a zoom-in included to make observational uncertainties visible.}
    \label{fig:Hubble banana}
\end{figure*}

The Hubble tension provides an important motivation to consider simulations with enhanced structure formation on large scales compared to $\Lambda$CDM. This is because the increased cosmic variance in the local $cz'$ would reduce the statistical significance of the mismatch between it and the expected background $\dot{a}$ from CMB observations. Bearing this in mind, we use \autoref{fig:Hubble banana} to show the apparent local Hubble and deceleration parameters (Section~\ref{sec: Hubble constant}). These results can be compared with the observed values from \citet{Camarena_2020a}, which we show towards the bottom right of each panel as a point with uncertainties. All simulations struggle to match the observations, but $\nu$HDM comes closest due to the very wide range of apparent deceleration parameters that it produces. This wide range makes it unlikely that the observed value should be so close to the background value of $-0.53$. We attribute this issue to the late and rapid growth of structure in $\nu$HDM. Our results suggest that the observed local underdensity would have built up much more gradually over most of the Hubble time, whereas in $\nu$HDM, structure formation is fairly slow until accelerations enter the MOND regime at $z \la 50$.

To gain further insight, we colour the result of each VP in \autoref{fig:Hubble banana} according to the apparent density contrast $\delta_{\rm app}$. VPs with a similar $\delta_{\rm app}$ define a similar track in the plane of apparent Hubble and deceleration parameters. VPs with a different $\delta_{\rm app}$ define a parallel track. Interestingly, the $\nu$HDM track that passes close to the observed point corresponds to $\delta_{\rm app} \approx -0.5$, though given the uncertainties, a somewhat smaller underdensity also seems to fit. The $\nu$HDM model therefore provides a reasonable fit to the observations of \citet{Camarena_2020a}. This success should be replicated in a different model that enhances structure formation over $\Lambda$CDM to a far smaller extent given $\nu$HDM produces too much structure (\autoref{fig:Mean bulk flow}). Out of the models we consider, those based on GR are not capable of achieving this, but those based on MOND come fairly close, with both $\nu$CDM and $\nu$HDM capable of solving the Hubble tension.

\subsection{Hubble dipole}
\label{sec:Results Hubble dipole}

The velocity field in the local Universe can also be quantified in terms of the apparent dipole in $H_0$ (Section~\ref{sec: Hubble dipole}). A detection was claimed by \citet{Migkas_2021} using galaxy clusters, whose relative distances can supposedly be found from various scaling relations. They use galaxy clusters across a range of redshifts, with the typical redshift corresponding to a distance of about $350/h$~Mpc. The Hubble dipole (direction of highest $H_0$) is reported to lie towards $l = 93^{\circ}$, $b  = 11^{\circ}$ in Galactic coordinates. This is almost opposite to the bulk flow of galaxies on smaller scales \citep[$l = 298^{\circ}$, $b = -8^{\circ}$;][]{Watkins_2023}. Since redshifts are inflated parallel to a bulk flow, we would expect the Hubble dipole to roughly align with the bulk flow on slightly smaller scales. Instead, the dot product between the above two directions is $\cos \theta_{\rm obs} = -0.91$, indicating almost completely the opposite.

\begin{figure*}
    \centering
    \includegraphics[width=\textwidth]{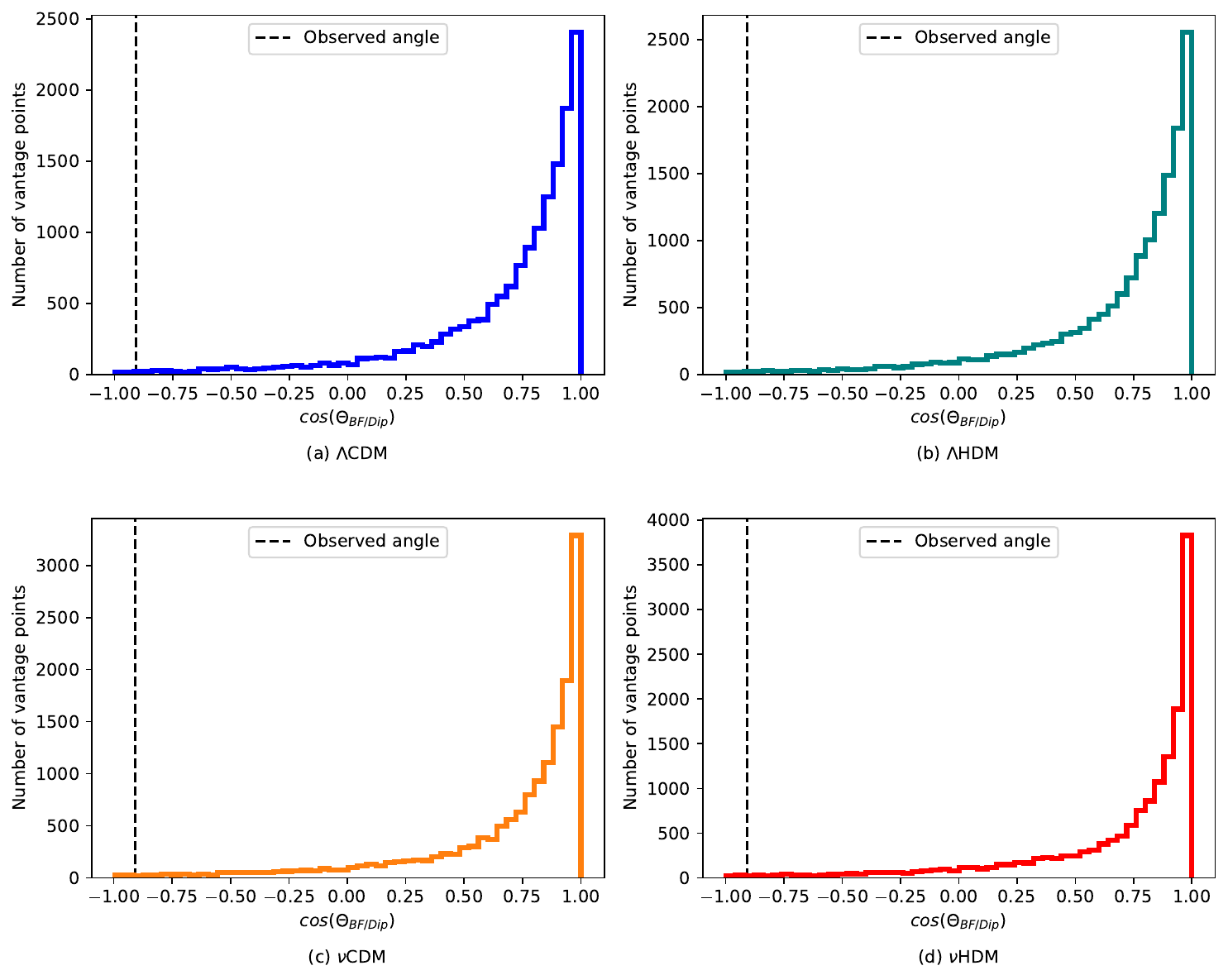}
    \caption{Histograms showing $\cos \theta$ between the bulk flow out to $200/h$~Mpc and the apparent Hubble dipole at $350/h$~Mpc in our simulation of $\Lambda$CDM (top left), $\Lambda$HDM (top right), $\nu$CDM (bottom left), and $\nu$HDM (bottom right). $\cos \theta = +1$ indicates alignment, while $-1$ indicates anti-alignment. The black dashed line is the observed angle \citep{Migkas_2021, Watkins_2023}. This is rarely reproduced in any simulation.}
    \label{fig:Theta hist}
\end{figure*}

To quantify the likelihood of $\cos \theta \approx -1$ in our simulations, we use \autoref{fig:Theta hist} to show $\cos \theta$ between the bulk flow out to $200/h$~Mpc and the Hubble dipole out to $350/h$~Mpc. Our results confirm that $\cos \theta$ is generally close to $+1$. A larger angle than reported observationally (lower $\cos \theta$) is very rare, arising about 0.3\% of the time ($2.75\sigma$ tension). This result is fairly consistent across all considered simulations, presumably because $\cos \theta$ is a function of rather large-scale modes in the initial conditions, which are similar between the different models.

\begin{figure*}
    \centering
    \includegraphics[width=\textwidth]{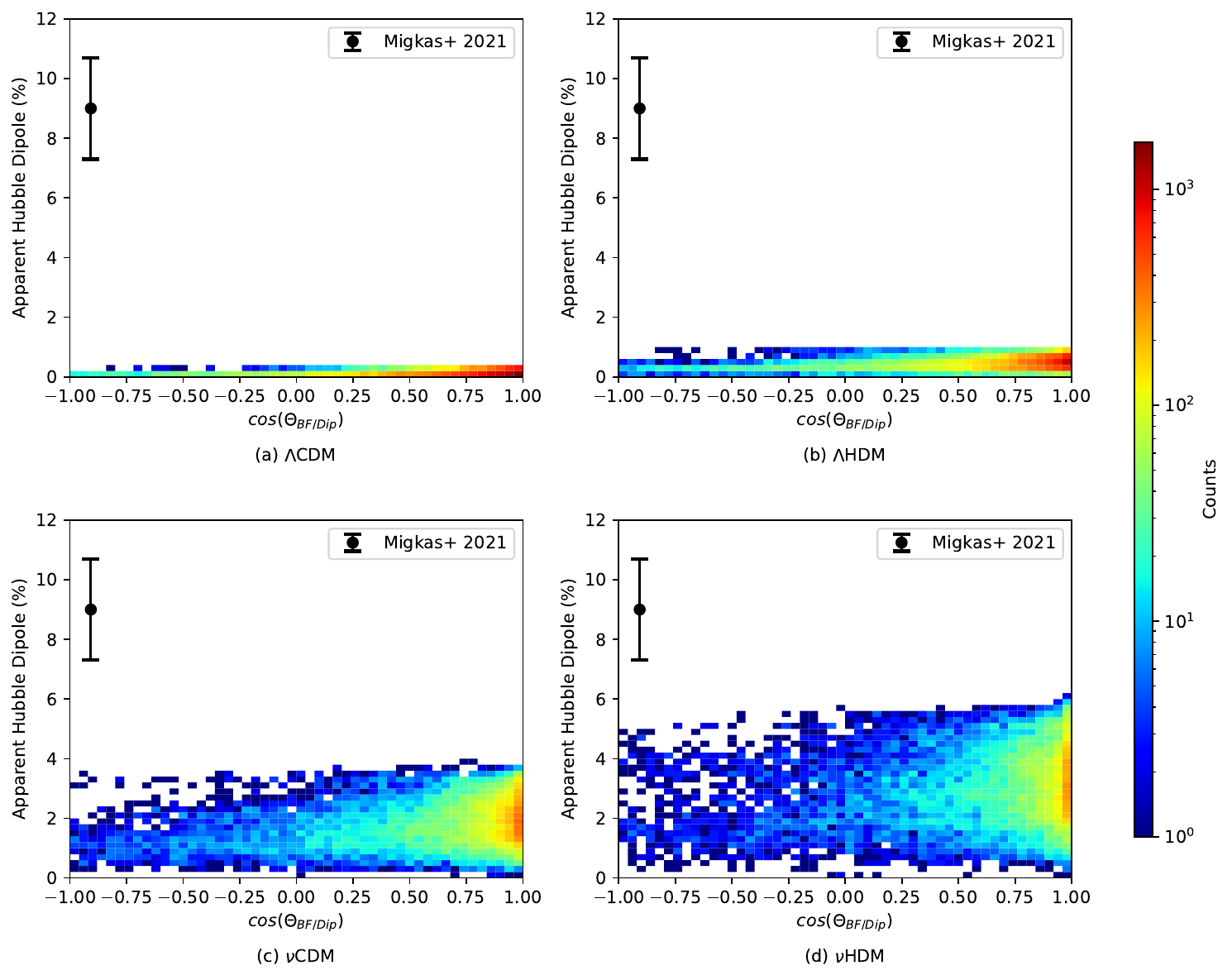}
    \caption{Heatmaps of the apparent Hubble dipole and the angle between it and the bulk flow in our simulation of $\Lambda$CDM (top left), $\Lambda$HDM (top right), $\nu$CDM (bottom left), and $\nu$HDM (bottom right). The black point with error bars shows the observed quantities \citep{Migkas_2021, Watkins_2023}. Changing CDM to HDM and changing from GR to MOND both increase the maximum apparent Hubble dipole. Even so, none of the simulations are able to reproduce the magnitude of the Hubble dipole reported by \citet{Migkas_2021}. Their reported anti-alignment between the Hubble dipole and bulk flow curve is also quite rare in our simulations.}
    \label{fig:Theta_vs_dipole}
\end{figure*}

The reported dipole amplitude of $9.0 \pm 1.7\%$ is also very rare in our simulations, as seen from the joint distribution of $\cos \theta$ and the percentage Hubble dipole (\autoref{fig:Theta_vs_dipole}). The $\nu$HDM simulation produces the largest dipoles, but it almost never yields dipoles $\ga 5\%$, not even half the value reported by \citet{Migkas_2021}.

Our results suggest that their claimed Hubble dipole is unlikely to be sourced by a large-scale bulk flow. The flip in direction and large amplitude at such a large distance appear to be rare in models with initial conditions and a background cosmology similar to those outlined by \citet{Planck_2020}. Moreover, the claimed dipole of \citet{Migkas_2021} is in the opposite direction to several other studies \citep[][and references therein]{Peebles_2022}, suggesting that the problem may lie with the novel use of galaxy cluster scaling relations as distance indicators \citep{Migkas_2021}. A recent detailed analysis has found that there is no statistically significant evidence for anisotropy in these scaling relations, limiting the magnitude of anisotropy in the local $cz'$ to $<3.2\%$ and the corresponding bulk flow amplitude to $<1300$~km/s \citep{Yasin_2026}.

\section{Discussion}
\label{sec:Discussion}

Our results in \autoref{fig:CHMF} indicate that the MOND cosmologies we consider produce far too much structure in a Hubble time, as seen in earlier simulations \citep{Angus_2011_clusters, Angus_2013}. This contradicts expectations from \citetalias{Haslbauer_2020} that combining HDM with MOND gravity would give approximately the correct amount of structure. Instead, the $\nu$HDM paradigm appears to suffer from the flaws of both $\Lambda$CDM and MOND in a `worst of both worlds' scenario -- galaxies form far later due to the lack of CDM \citep[Section~\ref{sec:Late galaxy formation}; see also][]{Wittenburg_2023}, but there would be far too much structure on large scales by $z = 0$. Since MOND is not compatible with a large amount of CDM, the anomalies in galaxy clusters and the CMB most likely need to be explained using HDM \citep[][and references therein]{Banik_Zhao_2022}. This inevitably leads to a top-down structure formation scenario in which large structures form first and then fragment into smaller structures \citep*{Bode_2001}. Gravitational accelerations from structure in the universe become $\la a_{_0}$ when $z \la 50$ \citepalias[section~3.1.3 of][]{Haslbauer_2020}. This gives a vast amount of time between $z = 50$ and the present epoch for MOND to enhance the growth of structure. The end result is a massive overprediction of peculiar velocities, regardless of whether we use CDM or HDM (\autoref{fig:hists pec vel}). This invalidates simple hybrid models of MOND + DM, including $\nu$HDM. 

Our results should not be taken to favour the $\Lambda$CDM paradigm given the serious difficulties that it faces. In particular, there is the well-known Hubble tension, a $>5\sigma$ mismatch between the local $cz'$ and the predicted $\dot{a} = H_0^{\mathrm{Planck}}$ from calibrating $\Lambda$CDM to the CMB anisotropies \citep{H0DN_2026, Valentino_2025}. The local $cz'$ is most strongly constrained by the Type Ia supernovae distance ladder and seems to be insensitive to the distance calibration technique \citep{Uddin_2024} or improvements to the distance calibrations measured between the \textit{Hubble} and \textit{James Webb Space Telescopes} \citep{Riess_2024_consistency, Freedman_2025}. Likewise, improved measurements of the CMB temperature anisotropies from both space-based \citep{Planck_2020} and ground-based \citep{ACT_2025, Camphuis_2025} facilities have continued the tension, maintaining it as one of the most pressing issues in modern cosmology. The Hubble tension cannot be due to inaccurate measurements of the CMB anisotropies, since $\Lambda$CDM requires a low $H_0$ to explain the combination of the CMB monopole temperature, primordial light element abundances, and BAO data \citep{Banik_2025_cosmology}. Other issues include the observed properties of the KBC void \citep{Keenan_2013, Wong_2022} and the high redshift, mass, and collision velocity of the El Gordo interacting galaxy clusters, which contradicts $\Lambda$CDM for any plausible infall velocity \citep*{Asencio_2021, Asencio_2023}. The El Gordo and KBC void observations suggest a need to enhance the growth of structure on large scales \citepalias{Haslbauer_2020}, as also suggested by the anomalously fast bulk flows reported by \citet{Watkins_2023} and confirmed by \citet*{Whitford_2023}.

We can consider our simulations as giving some idea of how a cosmological model might behave if structure formation on large scales is indeed faster than in $\Lambda$CDM, in which case the Hubble tension could be solved by outflow from a local void \citep{Keenan_2016, Shanks_2019a, Shanks_2019b, Ding_2020, Martin_2023, Cai_2025}. Some VPs in our MOND models come within $1-2\sigma$ of the locally observed values of the Hubble and deceleration parameters, but these typically deviate from the true background values far too much, again highlighting the need for a model which does not depart from $\Lambda$CDM as drastically as the $\nu$HDM simulations considered here (\autoref{fig:Hubble banana}). Our results suggest that if a local void is to solve the Hubble tension, then the underdensity needs to grow steadily rather than rapidly accelerate when $z \la 50$, as happens in $\nu$HDM. In particular, this should lead to a deceleration parameter that deviates less from the background value, thereby agreeing better with the observations. Since our MOND models produce far too much structure, it is quite plausible that a different model which deviates from $\Lambda$CDM to a smaller extent is still able to form analogues to the KBC void frequently enough that it does not falsify the model. Indeed, the KBC void might well be not that common in the correct cosmological model. We note that the deviations from $\Lambda$CDM needed to form the KBC void must be associated with the growth of structure because its 300~cMpc scale is very well observed in the CMB power spectrum, where there is excellent agreement with the expected temperature fluctuations \citep[the corresponding multipole moment is about 500;][]{Planck_2020, Tristram_2024}.

\subsection{When galaxies form without CDM}
\label{sec:Late galaxy formation}

It has previously been argued that some form of DM is required even in MOND \citep{Angus_2009, Haslbauer_2020, Banik_Zhao_2022}. The tight RAR in rotating disc galaxies suggests that galaxies are purely baryonic and lack CDM \citep{Li_2018, Desmond_2023}. However, CDM is an important ingredient in standard galaxy formation theory \citep{White_Rees_78}. It is therefore important to explore whether galaxies can form in a realistic timeframe without CDM, but with the gravity law altered to MOND.

Simple analytic estimates prior to accurate CMB observations suggested that galaxies should form in MOND cosmologies by $z = 10$ \citep{Sanders_1998_cosmology, Sanders_2008_highz_gals}, in good agreement with recent observations of early massive galaxies observed by the \emph{JWST} \citep{McGaugh_2024}. However, detailed hydrodynamical cosmological $\nu$HDM simulations found that galaxies only form by $z = 4$ in this model \citep{Wittenburg_2023}. Those authors carefully considered the initial power spectrum, which is suppressed on the comoving scales of galaxies due to the lack of CDM (see their figure~2). We show below that this discrepancy is driven by differences in the respective background cosmologies.

For a spherical region of radius $\lambda_{c}$ with corresponding mass $M_{c}$, equation~7 of \citet{Sanders_2008_highz_gals} predicts that density fluctuations in the MOND regime during the matter-dominated era will grow to unity at a redshift 
\begin{equation}\label{eq:Unity redshift}
    z_{1}=\left(\frac{27}{4a_{0}}\Omega \lambda_{c} H_{0}^{2}\right)^{-1/2} - 1,
\end{equation}
where $\Omega$ is the density parameter of the collapsing matter. $\lambda_{c}$ and $M_{c}$ are related by
\begin{equation}
    \lambda_{c} = \sqrt[3]{\frac{3M_{c}}{4\pi\rho_{0}}},
\end{equation}
where $\rho_{0}$ is the background density of the collapsing matter at $z=0$. \autoref{eq:Unity redshift} can therefore be further generalized as
\begin{equation}\label{eq:Sanders redshift}
    z_{1} = \left(\frac{27}{4a_{0}}\right)^{-1/2} \left(2GH_{0}^{4} \right)^{-1/6} \left(M_{c}\right)^{-1/6} \left( \Omega \right)^{-1/3} - 1.
\end{equation}
Interestingly, this formula is largely independent of the initial fluctuation amplitude. This is visualized in Figure 1 of \citet{Sanders_2008_highz_gals}, which states that ``MOND drives structure growth towards $(1+z)^{-2}$ quite independently of the initial conditions.'' They also used a spherically symmetric $N$-body code to show that these perturbations collapse to form galaxies within $\approx 1$ Gyr of reaching unity.

To obtain these promising results, \citet{Sanders_2008_highz_gals} assumed a flat FRW cosmology without dark matter such that $\Omega_{dm}=0$, $\Omega_{b}=0.04$, and $\Omega_{\Lambda}=0.96$ for, respectively, the dark matter, baryon, and dark energy density in units of the critical density, with a low present expansion rate of $H_0 = 63$~km/s/Mpc. They assumed a MOND acceleration parameter of $a_{0} = 1\times 10^{-10}\mbox{m/s}^{2}$, very similar to the modern value. With these parameters (and setting $\Omega=\Omega_{b}$), a benchmark perturbation mass of $M_{c}=10^{11} \, M_{\odot}$ will reach unity at $z \approx 40$ according to \autoref{eq:Sanders redshift}. The corresponding redshift of galaxy formation $\approx 1$ Gyr later is $z \ga 10$.

We can now compare how a density perturbation of the same mass evolves within this framework for the $\nu$HDM cosmology simulated by \citet{Wittenburg_2023}. We will assume that density perturbations on galactic scales are baryonic only. Free-streaming of the HDM prevents the growth of perturbations of this component on galactic scales by design -- this effect can be seen in figure~10 of \citet{Wittenburg_2023}, where structures below a mass of $10^{12} \, M_{\odot}$ contain virtually no HDM. This approach is theoretically valid as \autoref{eq:Unity redshift} assumes MOND effects apply to perturbations with respect to the background density, which only affects the growth of perturbations through its contribution to the background expansion rate.

The cosmological parameters in the $\nu$HDM model are taken directly from \citet{Planck_2020}, i.e., $\Omega_{dm}=0.266$, $\Omega_{b}=0.049$, $\Omega_{\Lambda}=0.685$, and $H_0 = 67.4$~km/s/Mpc. The MOND acceleration parameter is also slightly different at $a_0 =1.2\times 10^{-10}\mbox{m/s}^{2}$. Setting $\Omega=\Omega_{b}$ (since only baryonic perturbations will collapse on galactic scales) and again using a benchmark perturbation mass of $M_{c} = 10^{11} \, M_{\odot}$ gives a similar redshift of $z \approx 40$ for the density contrast to reach order unity. However, the higher $\Omega_{m}$ in this model leads to a higher $dz/dt$ at high $z$, so the redshift of galaxy formation $\approx 1$~Gyr later is now $z \ga 5.4$. This is more in line with the redshift of the first galaxies reported by \citet{Wittenburg_2023} in their hydrodynamical cosmological $\nu$HDM simulations.

Comparing the age of the universe between both models highlights this difference. In the \citet{Sanders_2008_highz_gals} cosmology, the universe is $\approx 0.18$ Gyr old at $z=40$ and $\approx 1.36$ Gyr old at $z=10$. Meanwhile, the \citet{Wittenburg_2023} cosmology is only $\approx 0.06$ Gyr old at $z=40$ and $\approx 0.47$ Gyr old at $z=10$. This would leave only 400~Myr for order unity perturbations to collapse, dynamically relax, and then form galaxies by $z=10$.

We note here that the redshift when the first galaxies form in $\nu$HDM \citep[$\approx5.4$ analytically and $\approx4$ from ][]{Wittenburg_2023} are both approximate and would require more detailed calculations to confirm. The analytic estimate is likely an upper bound on redshift. While density perturbations with the same $\lambda_{c}$ will converge to the same amplitude at a given redshift regardless of their initial amplitude, this convergence is not instantaneous and occurs later for perturbations with lower initial amplitude \citep[see figure~1 of ][]{Sanders_2008_highz_gals}. The time for perturbations to collapse and virialize after reaching unity should still be $\approx1$ Gyr as the calculations by \citet{Sanders_2008_highz_gals} assume the collapsing region has decoupled from the background cosmology. We note that a mass of $10^{11} M_{\odot}$ lies near the resolution of our simulations. The halo finder used also assumes Newtonian gravity, so the earliest structures were likely gravitationally bound before $z=4$. Even so, it is difficult to understand how galaxies can form by $z = 14$.

While beyond the scope of this work, a more involved application of the \citet{Sanders_2008_highz_gals} methodology to the $\nu$HDM framework would allow for a more conclusive test, but we can expect the true value to lie approximately in the range $4 \la z \la 5.5$ with a \citet{Planck_2020} background cosmology. It is clear both theoretically and through simulations that the epoch of galaxy formation depends on the background cosmology for these MOND models. The prediction that galaxies should form by $z=10$ \textit{is not a generic prediction of all cosmologies with MOND}; this redshift is sensitive to the underlying cosmological parameters. The above discussion highlights that a major reason for the high redshift at which galaxies are expected to form in the \citet{Sanders_2008_highz_gals} cosmology is that it uses an outdated background cosmology with a far older universe at $z = 10$.

In the case of $\nu$HDM, optimisations to the background cosmology are unlikely to solve this issue. While the cosmologies considered here are based on the \citet{Planck_2020} parameters for simplicity, \citet{Samaras_2025} find that a flat ``opt-$\nu$HDM'' cosmology with $H_{0} \approx 55$ km/s/Mpc, $\Omega_{dm} \approx 0.41$, $\Omega_{b}\approx0.09$, and $\Omega_{\Lambda} \approx 0.5$ provides the optimal fit to the CMB power spectrum for the $\nu$HDM model. The high $\Omega_{\mathrm{m}}$ is in $>20\sigma$ disagreement with BAO results from the Dark Energy Spectroscopic Instrument Data Release 2 (DESI DR2), which shows that $\Omega_{\mathrm{m}} = 0.2975 \pm 0.0086$ without combining with any external datasets \citep{DESI_2025}. It also provides no tangible improvement to when structures form. A $10^{11} \, M_{\odot}$ perturbation would reach unity at $z\approx37$ and would collapse and virialize $\approx1$~Gyr later, when $z\approx5$.

Low $\Omega_{m}$ models also face other issues, the most obvious being cosmology-independent constraints on the age of the Universe from the ages of the oldest Galactic stars and globular clusters. The \citet{Sanders_2008_highz_gals} model has an age of 24~Gyr at $z=0$. When the first galaxies form at $z\approx10$, the age is~1.35 Gyr. The oldest stars in such a model would therefore be $\approx20$~Gyr old. However, \citet{Cimatti_2023} and \citet{Tomasetti_2026} find stellar ages in the Milky Way consistent with an $\approx14$~Gyr old Universe once contaminants have been removed \citep[see also][]{Xiang_2022, Valcin_2021, Valcin_2025, Xiang_2025}. The high third peak of the CMB temperature anisotropies may also be problematic for such models due to the lack of dark matter, but this issue may be alleviated in newer relativistic MOND formulations \citep[e.g.][]{Skordis_2021, Skordis_2022, Blanchet_2024, Durakovic_2024}.

The problem of late galaxy formation will likely be faced by any cosmological model where MOND gravity is applied to density perturbations in an FRW background given the inevitable lack of CDM. For $\nu$HDM specifically, structure is formed too rapidly and too recently, producing too little at early times and too much at late times. The available hydrodynamical cosmological $\nu$HDM simulations indicate that MOND with DM does not produce the high redshift galaxies observed by the \emph{JWST}. This reveals a fundamental problem with the $\nu$HDM paradigm, invalidating it as a viable theory of structure formation.

\subsection{Particle physics and cosmology constraints on sterile neutrinos}
\label{sec:Neutrino masses}

In our simulations and those of \citet{Wittenburg_2023}, the HDM component in $\nu$HDM is assumed to be 11~eV sterile neutrinos, as proposed by \citet{Angus_2009}. None of the ordinary neutrinos can play the role of HDM as their rest energy is at most 0.45~eV \citep{Katrin_2019, Katrin_2022, Katrin_2025}. Sterile neutrinos as dark matter was first proposed by \citet{Dodelson_1994}. They are an attractive candidate as they are physically motivated by the standard model of particle physics and are expected to be weakly interacting. However, recent particle physics experiments cast doubt on the existence of sterile neutrinos \citep{KATRIN_2025_sterile, MicroBooNE_2025}. While the HDM in $\nu$HDM does not necessarily have to be 11~eV sterile neutrinos, the physical motivation for the model is weakened by these null detections.

Another problem with the 11~eV sterile neutrino hypothesis is that these would contribute an extra relativistic degree of freedom during the BBN era. This would increase the expansion rate, reducing the time between any two fixed temperature thresholds and thus the decay of free neutrons, increasing the primordial helium abundance \citep{Cyburt_2016}. Recent results indicate that the effective number of neutrino species during the BBN era was close to the standard value of 3, with 4 excluded at quite high confidence \citep{Kirilova_2024}. Since the sterile neutrinos are expected to be thermalised, this is an important issue for the $\nu$HDM paradigm. It could be avoided if the sterile neutrinos were more massive and not fully thermalised during the BBN era, but too high a mass would start to affect galaxies, undermining the MOND rotation curve fits \citep{Angus_2010_minimum_neutrino_mass}.

\subsection{Constraints on MOND from bound systems}
\label{sec:Discussion MOND}

Our simulations show that naively applying MOND to large scales using the Jeans swindle approach \citep{Nusser_2002} and adding HDM to fit galaxy clusters and the CMB leads to results incompatible with the observed large-scale structure. Our results are in line with difficulties faced by MOND in the outskirts of galaxy clusters -- even if we were to add the required amount of DM to fit the kinematics in their central regions \citep{Li_2023}. This is also evident in figure~5 of \citet{Kelleher_2024}, where the inferred $g$ at low accelerations (large radii) falls below the MOND expectation of $\sqrt{g_{_{\mathrm{N}}} a_{_0}}$.\footnote{The clusters A644 and A2319 should not be considered due to ongoing merger signatures.} The results instead suggest that $g_{_{\mathrm{N}}}$ should be enhanced by a fixed factor, which interestingly is similar to $\Omega_{\mathrm{m}}/\Omega_b$ inferred in $\Lambda$CDM using the CMB anisotropies \citep{Planck_2020, Tristram_2024}. This is to be expected if gravity follows the Newtonian inverse square law in the low acceleration regime and the ratio of enclosed baryonic to total matter in galaxy clusters is similar to that in the universe as a whole by the time we get to their outskirts. If a viable MOND + dark matter cosmological model exists, we would need to suppress MOND effects on large scales despite the low accelerations, in order to fit the cluster RAR and the observed CHMF and large-scale structure.

In addition to difficulties faced by MOND in the outskirts of galaxy clusters, it also struggles to match Solar System ephemerides, with Cassini radio tracking data from Saturn orbit falsifying the classical AQUAL and QUMOND modified gravity formulations of MOND at $8.7\sigma$ confidence in combination with constraints on $a_{_0}$ and the interpolating function from galaxy rotation curves \citep*{Desmond_2024}. The orbital energy distribution of long-period comets and the inclination distribution of Kuiper belt objects have also been used to rule out the AQUAL formulation of MOND \citep*{Vokrouhlicky_2024}. In particular, the Oort Cloud would not be stable on Gyr timescales in MOND gravity and Kuiper Belt objects would have a much broader range of inclinations to the Ecliptic than is observed, suggesting that MOND effects should be suppressed in the outer Solar System.

There is ongoing controversy on whether the dynamics of wide binary stars in the Solar neighbourhood are consistent with MOND. The groups of \citet{Pittordis_2023} and \citet{Banik_2024_WBT} claim that wide binaries favour Newtonian gravity, in the latter case apparently ruling out MOND at $16\sigma$ confidence. However, other studies instead claim that wide binaries favour MOND with an external field effect \citep{Hernandez_2022, Hernandez_2023, Hernandez_2024_statistical, Chae_2023, Chae_2024a, Chae_2024b}. For instance, \citet{Chae_2024b} found a $6 - 9\sigma$ preference for MOND. A review by \citet{Hernandez_review_2024} has highlighted issues in \citet{Pittordis_2023} and \citet{Banik_2024_WBT} that may have erroneously biased their results towards Newtonian gravity. Of particular concern are the fitting of noise free templates to noisy data in \citet{Pittordis_2023}, uncertainties that are larger than data bin widths in \citet{Banik_2024_WBT}, the lack of a `Newtonian anchor' regime as a reference in both, and the modelling of hidden tertiary companions. For example, the tertiary companion fraction in \citet{Banik_2024_WBT} is inferred to be $63 \pm 1\%$ in their nominal analysis (see their table~3), whereas more direct observations find it should be below $50\%$ \citep[e.g., see figure~39 of][]{Moe_2017}. We note that \citet{Banik_2024_WBT} argued in their section~5.1.3 that further refinements to the unknown distribution of close binary orbital radii could reduce the close binary fraction to perhaps 50\%, but even so, the large fraction of close binaries makes it difficult to be sure of the gravity law. There are good prospects for better characterising the close binary population in the near future \citep*{Manchanda_2023}. This field still appears to be under debate, e.g. \citet{Pittordis_2025} address the criticisms of their earlier work and still maintain preference for Newtonian gravity, while \citet{Chae_2026} apply new data from HARPS and again finds preference for MOND. We also note the appearance of three recent studies that argue local wide binaries favour Newton over MOND \citep{Saad_2025, Makarov_2026, Cookson_2026}, with \citet{Cookson_2026} suggesting that earlier claims in favour of MOND were driven by selection effects. 

Despite some issues and debate on scales smaller and larger than galaxies, MOND is still successful with the internal dynamics of galaxies, including with regards to the tidal stability of dwarf galaxies in cluster environments \citep{Asencio_2022} and the origin of the planes of satellite galaxies in the LG from a past flyby encounter between the Milky Way and Andromeda galaxies \citep{Banik_Ryan_2018, Bilek_2018, Banik_2022_satellite_plane}. This flyby is inevitable in MOND \citep{Zhao_2013} and may be testable in future through anisotropy of distant Galactic tidal streams \citep*{Asencio_2025}. The asymmetric tidal tails of open star clusters in the Solar neighbourhood have also been argued to favour MOND \citep{Kroupa_2024}. These results extend the success of MOND on galaxy scales beyond equilibrium dynamics, with the external field effect playing a crucial role in tidal stability considerations. Galaxy-galaxy weak lensing by isolated lens galaxies down to $g_{_{\mathrm{N}}} \approx 10^{-15}$~m/s\textsuperscript{2} or about $10^{-5} \, a_{_0}$ also seems to follow MOND predictions \citep{Brouwer_2021, Mistele_2024_flatRC, Mistele_2024_RAR}. It is not presently clear how these results can be understood without the action of a MOND-like force law.

\section{Conclusions}
\label{sec:Conclusions}

In this work, we present the largest collisionless $N$-body simulations of a MOND cosmology to date, with a box size of $800/h$~cMpc and $256^3$ particles. These simulations aimed to compare the $\Lambda$CDM standard model of cosmology to the newer $\nu$HDM model \citep{Angus_2009}, which combines MOND gravity ($\nu$) with HDM in the form of sterile neutrinos with a rest energy of 11~eV. The resulting cosmology is assumed to have an identical expansion rate history to $\Lambda$CDM and adequately fits the CMB power spectrum almost as well as $\Lambda$CDM \citep[figure~2 of][]{Wittenburg_2023}. The model was also expected to reproduce galaxy cluster dynamics thanks to the HDM component \citep{Angus_2011_clusters} and to reproduce MOND fits to galaxy rotation curves given the low density of HDM \citep{Angus_2010}. $\nu$HDM has previously been simulated in smaller boxes of size 256 and 512~Mpc/$h$ \citep{Angus_2011_clusters, Angus_2013}. For further discussion of the $\nu$HDM paradigm, we refer the reader to section~3.1 of \citetalias{Haslbauer_2020}.

We performed simulations of four distinct cosmological models: $\Lambda$CDM as a baseline, $\nu$HDM as our physically motivated alternate model, and two unphysical models we dubbed $\Lambda$HDM and $\nu$CDM to test the individual effects of using HDM or MOND gravity, respectively. The simulations were set up in a similar manner to that outlined in \citet{Wittenburg_2023}, with a \citet{Planck_2020} background cosmology. We disabled the hydrodynamical features for our runs, allowing a complementary study where we focus on cosmological aspects of the model in a large simulation box.

We find that the HDM models underproduce light haloes and overproduce massive haloes when compared to their CDM counterparts, consistent with previous simulations \citep{Angus_2013}. We also find that the MOND models produce structures at least $10\times$ more massive than their Newtonian counterparts. Our results build on \citet{Angus_2011_clusters} and \citet{Angus_2013} by extending this measurement to larger masses and comparing with updated measurements of the CHMF. The MOND models fail to reproduce observed halo counts from \citet{Driver_2022} at the high mass end, where haloes up to $5 \times10^{17} \, M_{\odot}/h$ form in $\nu$HDM by $z = 0$ (\autoref{fig:CHMF}). This general overproduction of structure is also evident in \autoref{fig:hists pec vel}, where the typical peculiar velocities of around 5000~km/s in $\nu$HDM leave the LG peculiar velocity of `only' 627~km/s as a $3.2\sigma$ outlier.

We also investigated the formation of supervoids analogous to the observed KBC void \citep*{Keenan_2013} or Local Hole \citep{Wong_2022}. Such phenomena were already seen to be produced in the $\nu$HDM model \citep{Angus_2013} and could potentially solve the Hubble tension, as outflows from a local supervoid could lead to an overestimation of the locally determined Hubble constant. Such supervoids are readily apparent in $\nu$HDM (\autoref{fig:Density maps}). When measuring the density contrast within $225/h$~Mpc of an observer, we confirm quantitively that structures analogous to the Local Hole can form in $\nu$HDM (\autoref{fig:hists delta}). No such analogous underdensities are found in either of the Newtonian simulations.

We then explored the effect of the local density contrast on the locally determined Hubble constant and deceleration parameter (\autoref{fig:Hubble banana}). We find all our simulations struggle to exactly match the observed values \citep{Camarena_2020a}. In the Newtonian simulations, this is because large enough values of $\mathcal{H}_{0}$ are not reached, whereas in the MOND models, the corresponding $\mathcal{Q}_{0}$ values are too negative. The overproduction of structure in the MOND models is once again evident here in that some regions of our $\nu$HDM simulations have an extreme value of $\mathcal{Q}_{0} \ge 20$. We also see evidence of the resulting supervoids overcorrecting the Hubble tension, with values ranging between $40<\mathcal{H}_{0}<100$~km/s/Mpc.

We found the MOND models also overcorrect the bulk flow tension (\autoref{fig:Mean bulk flow}). Whereas $\Lambda$CDM is ruled out at $>5\sigma$ confidence due to it \emph{underpredicting} the observed local bulk flow \citep{Watkins_2023}, we conversely find $\nu$HDM is ruled out to $>5\sigma$ confidence by \emph{overpredicting} the bulk flow curve. Ironically, it is only our `unphysical' $\Lambda$HDM model that appears to match this observation well, although this is likely a fluke simply due to the higher peculiar velocity field present in this model compared to $\Lambda$CDM.

The final cosmological tension we explored was the Hubble dipole observed by \citet{Migkas_2021}. If this dipole is sourced by some large scale bulk flow, it must have reversed direction compared to the bulk flow measured by \citet{Watkins_2023} on a slightly smaller scale. We find such a reversal to be rare across all of our simulations, with the dipole magnitude also being unreproducible if sourced by a bulk flow. We therefore suggest that a bulk flow origin for the dipole should be dismissed as it would evidently require a cosmological model more extreme than even the $\nu$HDM model. Regardless of the cosmological model, it is important to understand why the Hubble dipole reported by \citet{Migkas_2021} is in the opposite direction to most other studies \citep{Peebles_2022}.

We also discussed when the first galaxies are predicted to form in $\nu$HDM using the analytic formulae of \citet{Sanders_2008_highz_gals}. We found that with a flat background cosmology in which $\Omega_{m}\approx0.3$ and $H_{0}\approx70$ km/s/Mpc, the first galaxies are not expected to form before $z=5$ in a $\nu$HDM-like model. This prediction is consistent with the formation of the first galaxies at $z\approx4$ in the hydrodynamical $\nu$HDM simulations of \citet{Wittenburg_2023}. Such models will therefore be unable to form the high redshift galaxies observed by \emph{JWST} at $z \ga 14$ \citep{Carniani_2024}. The main reason for \citet{Sanders_2008_highz_gals} predicting such high redshift galaxies appears to be the use of an outdated background cosmology in which there is much more time at $z > 14$, but this also leads to a far older universe at $z = 0$, in tension with the ages of the oldest Galactic stars and globular clusters. A generic consequence of MOND is an absence of CDM, which can make it very difficult to form galaxies at high redshift.

Taking these results as a whole, we argue that $\nu$HDM, or any derivative thereof, can be discounted as a viable alternative cosmological model. We find the tension at the high mass end of the CHMF is worsened as the observations probe to higher masses \citep{Driver_2022} and that the typical peculiar velocities are an order of magnitude too high. The model also fails to solve key cosmological tensions for $\Lambda$CDM. The bulk flow tension is overcorrected, allowing us to reject $\nu$HDM at $>5\sigma$ confidence based on this alone. While the model does successfully produce analogues to the Local Hole \citep{Wong_2022}, this does not lead to a satisfactory supervoid solution to the Hubble tension. The observed Hubble dipole \citep{Migkas_2021} is also not reproduced. It appears that whenever the eventual solution(s) to these tensions is discovered, it will take the form of a much more subtle modification to the $\Lambda$CDM paradigm than the $\nu$HDM cosmology.

\section*{Acknowledgements}

The authors would like to thank the referee, whose comments have significantly improved the scientific content of this work. AR is supported by studentship ST/X508780/1 from the Science and Technology Facilities Council. IB is supported by Royal Society University Research Fellowship grant 211046. IB and HZ were supported by Science and Technology Facilities Council grant ST/V000861/1. OC was supported by a grant from the Student-Staff Council (SSC) and Physics Trust at the department of physics and astronomy at the University of Saint Andrews, as well as the Saint Andrews Research Internship Scheme (StARIS). HZ acknowledges the Fellowship for international cooperation from the University of Science and Technology of China, and a visiting professorship from Imperial College London. The authors are grateful to Stacy McGaugh and the HPC team at Case Western Reserve University for providing access to and aid with their computing cluster, on which the simulations were run. The authors would also like to thank Harry Desmond for helpful comments.

\section*{Data availability}

\citet{Nagesh_2021} provides a user guide for \textsc{por}, with the linked online repository containing the cosmological patch used in this contribution. The techniques used here are similar to \citet{Wittenburg_2023}, with the simplification that we do not consider baryonic physics processes. Halo catalogues can be made available upon reasonable request.

\bibliographystyle{mnras}
\bibliography{vHDM_bbl}


\bsp 
\label{lastpage}
\end{document}